\documentclass[aps,prb,amsmath,amssymb,twocolumn,superscriptaddress,floatfix,dvipsnames,longbibliography]{revtex4-2}
\usepackage{graphicx}
\usepackage{orcidlink}
\usepackage{hyperref}
\usepackage{dcolumn}
\usepackage[version=4]{mhchem}
\usepackage{tabularx, booktabs, siunitx, makecell}
\usepackage{bm}
\usepackage{float}
\hypersetup{pdftitle={Ktenasite},pdfauthor={Kaushick},pdfsubject={Frustrated magnetism},pdfdisplaydoctitle}

\def\ktD{Cu$_\text{2.7}$Zn$_\text{2.3}$(SO$_\text{4}$)$_\text{2}$(OD)$_\text{6}\cdot$6D$_\text{2}$O}
\def\ktH{Cu$_\text{2.7}$Zn$_\text{2.3}$(SO$_\text{4}$)$_\text{2}$(OH)$_\text{6}\cdot$6H$_\text{2}$O}
\def\TN{\ensuremath{T_\text{N}}}
\def\Cp{\ensuremath{c_P}}
\def\Cmag{\ensuremath{c_\text{mag}}}
\def\Smag{\ensuremath{\Delta S_\text{mag}}}
\begin{document}

\title{Disorder-driven magnetic duality in the spin-$\frac{1}{2}$ system ktenasite, \ktH}

\author{Kaushick K.\ Parui~\orcidlink{0000-0002-6584-7830}}
\email{kaushick.parui@tu-dresden.de}
\author{Anton A.\ Kulbakov~\orcidlink{0000-0001-9676-8310}}
%\email{anton.kulbakov@tu-dresden.de}
\affiliation{Institut f\"ur Festk\"orper- und Materialphysik, Technische Universit\"at Dresden, 01062 Dresden, Germany}

\author{Roman Gumeniuk~\orcidlink{0000-0002-5003-620X}}
%\email{roman.gumeniuk@physik.tu-freiberg.de}
\affiliation{Institut für Experimentelle Physik, TU Bergakademie Freiberg, 09596 Freiberg, Germany}

\author{Eduardo Carrillo-Aravena~\orcidlink{0009-0009-5293-7456}}
%\email{eduardo.carrillo@tu-dresden.de}
\affiliation{Fakult{\"a}t f{\"u}r Chemie und Lebensmittelchemie, Technische Universit{\"a}t Dresden, 01062 Dresden, Germany}
\affiliation{W\"urzburg-Dresden Cluster of Excellence on Complexity and Topology in Quantum Matter\,---\,ct.qmat, Technische Universit\"at Dresden, 01062 Dresden, Germany}

\author{Mar\'ia Teresa Fern\'andez-D\'iaz~\orcidlink{0009-0009-0524-0787}}
%\email{ferndiaz@ill.fr}
\affiliation{Institut Laue-Langevin, 71 avenue des Martyrs, CS 20156, 38042 Grenoble CEDEX 9, France}

\author{Stanislav Savvin~\orcidlink{0000-0002-2541-909X}}
%\email{savvin@ill.fr}
\affiliation{Institut Laue-Langevin, 71 avenue des Martyrs, CS 20156, 38042 Grenoble CEDEX 9, France}
\affiliation{Instituto de Nanociencia y Materiales de Arag\'{o}n, Facultad de Ciencias, CSIC--Universidad de Zaragoza, 50009 Zaragoza, Spain}

\author{Artem Korshunov~\orcidlink{0000-0001-5929-6473}}
%\email{artem.korshunov91@gmail.com}
\affiliation{Donostia International Physics Center (DIPC), Paseo Manuel de Lardiz\'abal, 20018 San Sebasti\'an, Spain}

\author{Sergey Granovsky~\orcidlink{0000-0001-9425-6696}}
%\email{sergey.granovsky@tu-dresden.de}
\affiliation{Institut f\"ur Festk\"orper- und Materialphysik, Technische Universit\"at Dresden, 01062 Dresden, Germany}

\author{Thomas Doert~\orcidlink{0000-0001-7523-9313}}
%\email{thomas.doert@tu-dresden.de}
\affiliation{Fakult{\"a}t f{\"u}r Chemie und Lebensmittelchemie, Technische Universit{\"a}t Dresden, 01062 Dresden, Germany}

\author{Dmytro S.\ Inosov~\orcidlink{0000-0003-0639-2503}}
\email{dmytro.inosov@tu-dresden.de}
\affiliation{Institut f\"ur Festk\"orper- und Materialphysik, Technische Universit\"at Dresden, 01062 Dresden, Germany}
\affiliation{W\"urzburg-Dresden Cluster of Excellence on Complexity and Topology in Quantum Matter\,---\,ct.qmat, Technische Universit\"at Dresden, 01062 Dresden, Germany}

\author{Darren C.\ Peets~\orcidlink{0000-0002-5456-574X}}
\email{darren.peets@tu-dresden.de}
\affiliation{Institut f\"ur Festk\"orper- und Materialphysik, Technische Universit\"at Dresden, 01062 Dresden, Germany}

\begin{abstract}
Disorder in frustrated quantum systems can critically influence their magnetic ground states and drive exotic correlated behavior. In the $S = \frac{1}{2}$ system ktenasite, \ktH, we show that structural disorder drives an unexpected dimensional crossover and stabilizes a rare coexistence of distinct magnetic states. Neutron diffraction reveals significant Cu/Zn mixing at the Cu2 site, which tunes the \ce{Cu^2+} sublattice from a two-dimensional scalene-distorted triangular lattice into a one-dimensional spin-chain network. Magnetic susceptibility, neutron diffraction, ac susceptibility, and specific heat measurements collectively indicate magnetic duality: a coexistence of incommensurate long-range magnetic order below $T_\text{N} = 4$\,K and a cluster spin-glass state with $T_\text{f} = 3.28$\,K at $\nu = 10$\,Hz. Our findings highlight ktenasite as a rare platform where structural disorder tunes the effective dimensionality and stabilizes coexisting ordered and glassy magnetic phases, offering a unique opportunity to explore the interplay of frustration, disorder, and dimensional crossover in quantum magnets. 

\end{abstract}
\maketitle 

\section{Introduction}
Disorder plays a critical role in condensed matter physics, profoundly influencing material properties and giving rise to diverse quantum phenomena. In doped semiconductors, controlled disorder via impurities enhances conductivity by introducing charge carriers. Conversely, in quantum magnets, disorder can destabilize long-range magnetic order (LRO). It manifests in various forms---chemical~\cite{Pal2018,Kulbakov2021, Dong2025} (e.g., site mixing or cation substitution such as Cu$^{2+}$/Zn$^{2+}$) and structural~\cite{Saha2023, Mannathanath_Chakkingal2025} (e.g., stacking faults, interstitials, or vacancies). Recent studies also highlight the role of proton disorder in modulating exchange interactions~\cite{Parui2025, Kulbakov2025b}, which, in conjunction with geometric frustration and competing interactions, can stabilize novel magnetic ground states.

The exploration of disorder-induced quantum phases is well established. Materials such as herbertsmithite~\cite{Shores2005}, YbMgGaO$_4$~\cite{Zhu2017}, Ba$_3$CuSb$_2$O$_9$~\cite{Smerald2015}, and Y$_2$CuTiO$_6$~\cite{Kundu2020a} exhibit spin-liquid mimicry driven by either substitutional disorder or intrinsic structural randomness. Similar spin-liquid–like behavior has also been reported in hydroxides such as \ce{CuSn(OH)6}, where proton disorder plays a crucial role in destabilizing LRO~\cite{Kulbakov2025a, Kulbakov2025b}. Various kinds of disorder, when coupled with frustration, may destabilize classical N\'eel order at low temperatures, promoting dynamic and exotic ground states.

Controlled substitution can suppress magnetic order, stabilizing spin-liquid-like or disordered glassy phases. For example, in spin-1/2 pyrochlore Lu$_2$Mo$_2$O$_7$, O$^{2-}$/N$^{3-}$ substitution yields a dynamic ground state~\cite{Clark2014,Uematsu2019}. In Ho$_2$Ti$_2$O$_7$~\cite{Savary2017}, Sc$^{3+}$ doping at Ti$^{4+}$ sites introduces oxygen vacancies, altering magnetic behavior. In Yb$_2$Ti$_2$O$_7$~\cite{Sala2018, Ghosh2018}, oxygen vacancies favor a quantum spin liquid, while Yb substitution on the Ti sites promotes ferromagnetism. Y$_2$CuTiO$_6$~\cite{Kundu2020a}, with 50:50 Cu$^{2+}$/Ti$^{4+}$ disorder on a triangular lattice, remains magnetically disordered down to 50\,mK, resembling a cooperative paramagnet. In Zn-doped averievite (Cu$_{5-x}$Zn$_x$V$_2$O$_{10}$CsCl)~\cite{Georgopoulou2023, Simutis2025}, increasing Zn content progressively disrupts LRO: while Cu$_5$VO$_{10}$CsCl exhibits a long-range magnetically ordered ground state, Cu$_4$ZnVO$_{10}$CsCl displays a spin-glass-like ground state, and Cu$_3$Zn$_2$V$_2$O$_{10}$CsCl supports a dynamic, fluctuating ground state indicative of kagome layer decoupling.

Ktenasite, (Cu,Zn)$_5$(SO$_4$)$_2$(OH)$_6 \cdot$6H$_2$O, is a rare copper-zinc hydroxysulfate mineral of the ktenasite group. Initially mischaracterized upon its discovery in 1950~\cite{Kokkoros1950}, its corrected composition~\cite{Raade1977,Mellini1978} reflects significant natural variation in the Cu/Zn ratio. It crystallizes in space group $P2_1/c$ with Cu and Zn occupying three distinct sites in the asymmetric unit. Zn-rich variants exhibit Zn substitution at both Cu1 and Cu2 sites. Synthetic powder samples have been prepared, wherein transmission electron microscopy (TEM) observations suggest the formation of a superlattice structure along the $a$-axis and pronounced crystal growth along the $b$ axis~\cite{Xue2004}. 

Here, we investigate synthetic ktenasite, which crystallizes in the $P2_1/c$ space group, where Cu$^{2+}$ ions occupy an anisotropic, scalene-distorted triangular lattice. However, intrinsic site disorder disrupts this frustrated geometry, promoting a dimensional reduction toward decoupled one-dimensional (1D) spin chains. Magnetization measurements reveal a broad hump accompanied by a bifurcation between ZFC and FC curves in ac susceptometry, which suggests a cluster spin-glass state, while the derivative of the magnetization additionally displays a sharp peak at $T_{\mathrm{N}} = 4$~K. The coexistence of both LRO and a glassy component is further evidenced by the presence of both a broad feature and a sharp $\lambda$-type anomaly in the specific heat at $T_{\mathrm{N}}$. Neutron diffraction reveals sharp magnetic Bragg reflections indicative of correlated, static magnetic moments. Thus, ktenasite exhibits a highly unusual coexistence of incommensurate LRO and cluster spin-glass behavior, despite\,---\,or perhaps because of\,---\,significant intrinsic disorder arising from Cu/Zn site mixing. This disorder not only drives a dimensional crossover from frustrated two-dimensional (2D) triangular layers to 1D chains but also promotes percolation within their spatially entangled network, stabilizing a dual magnetic ground state where disorder acts as a tuning parameter rather than a pertubation.

\section{Experimental Methods}

\ktH\ was prepared via a hydrothermal route under autogenous pressure. Initially, 10\,mL of 1\,M ZnSO$_4$ solution was prepared by adding 2.88\,g of Zn(SO$_\text{4}$)$\cdot$7H$_\text{2}$O (ThermoFisher GmbH, 99.0-103.0\%) into 10\,mL deionized water. This was followed by the addition of 1\,g \mbox{CuO} (ThermoFisher GmbH) nanopowder to the previous mixture. The resulting homogeneous mixture was subsequently transferred to a 50\,mL Teflon-lined stainless-steel autoclave, which was sealed and held at 40\,$^\circ$C in a furnace for 9 days. Following the heating phase, the autoclave was allowed to cool to room temperature naturally.  The resulting product was filtered, washed thoroughly with deionized water, then dried in a vacuum furnace at ambient temperature. The final product consisted of polycrystalline powder along with a few sub-millimeter-sized crystals of ktenasite, the latter requiring prolonged heating for formation (powders typically form after $\sim$\,4 days). A deuterated batch was prepared for neutron scattering experiments using D$_2$O (Acros Organics, 99.8\,at.\% D) in place of H$_2$O, aiming to reduce the incoherent scattering from hydrogen. To avoid loss of waters of crystallization, the samples were stored in a freezer as a precaution.

Differential scanning calorimetry (DSC) and thermal gravimetric analysis (TG) were performed on \ktH\ samples heated to 670\,K at a rate of 5 K min$^{-1}$ under argon atmosphere in an STA 409 C/CD simultaneous thermal analyzer (Netzsch GmbH \&\ Co. KG). 

The surface morphology of ktenasite powder was imaged using an in-lens detector on a Carl Zeiss AG\ Ultra 55 field-emission scanning electron microscope (SEM) at ambient temperature. The powder was sprinkled on a carbon-adhesive tape affixed to a sample puck. Energy-dispersive x-ray spectroscopy (EDS) was conducted with a Bruker\ Quantax EDS system, and data analysis was performed using Bruker's \textsc{Esprit} software package.  

A powder x-ray diffraction (PXRD) pattern was recorded at room temperature using a STOE Stadi P diffractometer in transmission mode. The instrument utilized Cu-K$\alpha_1$ radiation ($\lambda = 1.5406$\,\AA) to scan an angular range from 5.0$^\circ$ to 107.0$^\circ$ in $2\theta$.

Single-crystal x-ray diffraction (SCXRD) was measured on a Rigaku XtaLAB Synergy-S diffractometer equipped with a hybrid photon counting detector and a microfocus PhotonJet-S x-ray tube source using Mo-K$\alpha$ radiation ($\lambda = 0.7107$\,\AA) at 180\,K. Data were processed using \textsc{CrysAlisPro}, with Gaussian absorption corrections based on indexed crystal facets, followed by an empirical correction based on spherical harmonics. An initial model was created using \textsc{Shelxt}~\cite{Sheldrick2015a} and refined by full-matrix least-squares methods against $F_{\text{obs}}^{2}$ using \textsc{Shelxl}~\cite{Sheldrick2015b} within \textsc{Olex2}~\cite{Dolomanov2009}. All hydrogen positions were visible during differential Fourier synthesis in the later stages of the refinement and coincided with those observed in neutron diffraction. Hydrogen atoms from water molecules were refined using a riding model, whereas those from hydroxyl ions were refined freely. All non-hydrogen atoms were refined with anisotropic displacement parameters. The occupancies of the mixed Zn2/Cu2 site were determined from neutron diffraction data. Additional x-ray diffuse-scattering data were collected at the diffraction side station of the ID28 beamline at the ESRF in Grenoble, France\,\cite{ID28}. 

\begin{figure*}
  \includegraphics[width=\textwidth]{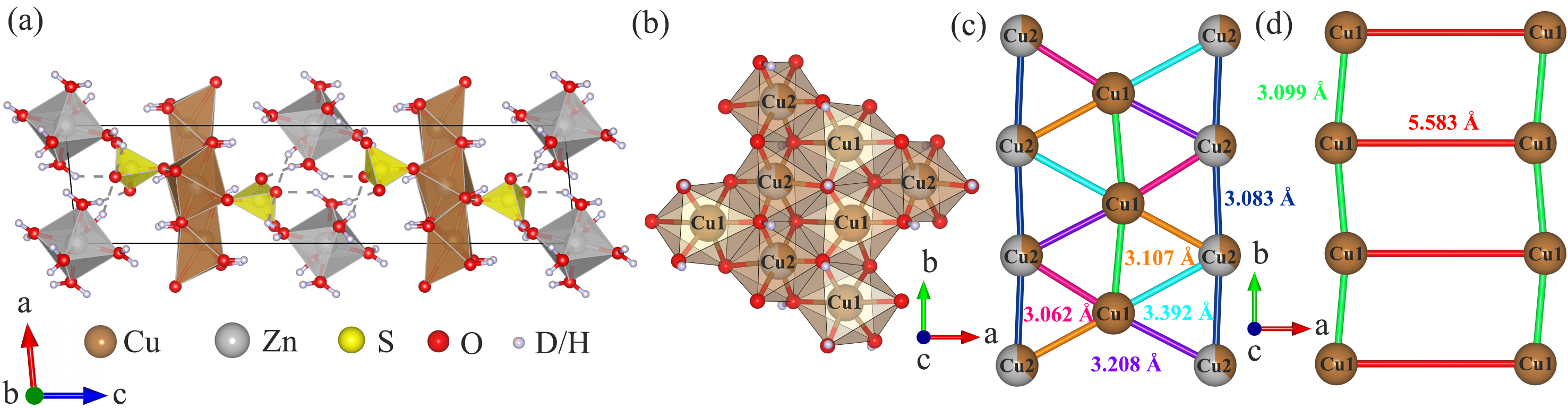}%\vspace{-10pt}
  \caption{\label{Crystal_Structure}(a) Refined crystal structure of \ktD\ in monoclinic $P2_1/c$ from D2B data at 10\,K. (b) Single $[\ce{Cu(Cu,Zn)(OD)3O}]^{2-}$ layer of edge-sharing, Jahn–Teller–distorted octahedra. (c,d) Magnetic sublattice showing (c) an anisotropic triangular arrangement of \ce{Cu^{2+}} ions, which evolves into (d) corrugated 1D chains along the $b$ axis at Cu1 sites due to random site disorder, with Cu–-Cu distances indicated.}
\end{figure*}

Neutron powder diffraction (NPD) patterns at 10\,K were collected over a $2\theta$ angular range from 8.0$^\circ$ to 160.0$^\circ$ at the Institut Laue-Langevin (ILL) in Grenoble, France, using the high-resolution two-axis diffractometer D2B~\cite{Hewat1986} with a wavelength $\lambda = 1.594$\,\AA. Additional measurements were performed on the high-intensity diffractometer XtremeD~\cite{Rodriguez-Velamazan2011} using 2.445-\AA\ neutrons at 1.5 and 11\,K. Both the powder XRD and NPD patterns were analyzed by the Rietveld refinement method~\cite{Rietveld1969} using the  \textsc{FullProf} software package~\cite{FullProf}. The crystal structure was visualized using {\sc Vesta}~\cite{VESTA}.

Temperature-dependent dc magnetization measurements were performed using a vibrating sample magnetometer (VSM) in a Cryogenic Ltd.\ Cryogen-Free Measurement System (CFMS). The sample, enclosed in a gelatin capsule within a plastic straw, was studied under zero-field-cooled-warming, field-cooled-cooling, and field-cooled-warming conditions. Isothermal magnetization was recorded at 2 and 50\,K in magnetic fields up to $\pm 14\,\text{T}$. The ac susceptibility was measured using an Oxford Instruments MagLab System2000 between 2.5 and 6\,K, with a 10-Oe ac field and frequencies ranging from 10 to 1000 Hz. 

Low-temperature specific-heat measurements were performed on a pressed pellet using the two-tau relaxation method in a Quantum Design Physical Property Measurement System (PPMS) DynaCool-12 system equipped with a $^3$He refrigerator. Addenda measurements were conducted beforehand to account for contributions from the sample holder and Apiezon N grease.
  
Fourier-transform infrared spectroscopy (FTIR) was measured using a Bruker Vertex 70 in attenuated total reflectance (ATR) construction between 400 and 4000\,cm$^{-1}$ with 2\,cm$^{-1}$ resolution on both \ktH\, and deuterated \ktD\, powders.  

\section{Crystal Structure}

\begin{figure}
  \includegraphics[width=\columnwidth]{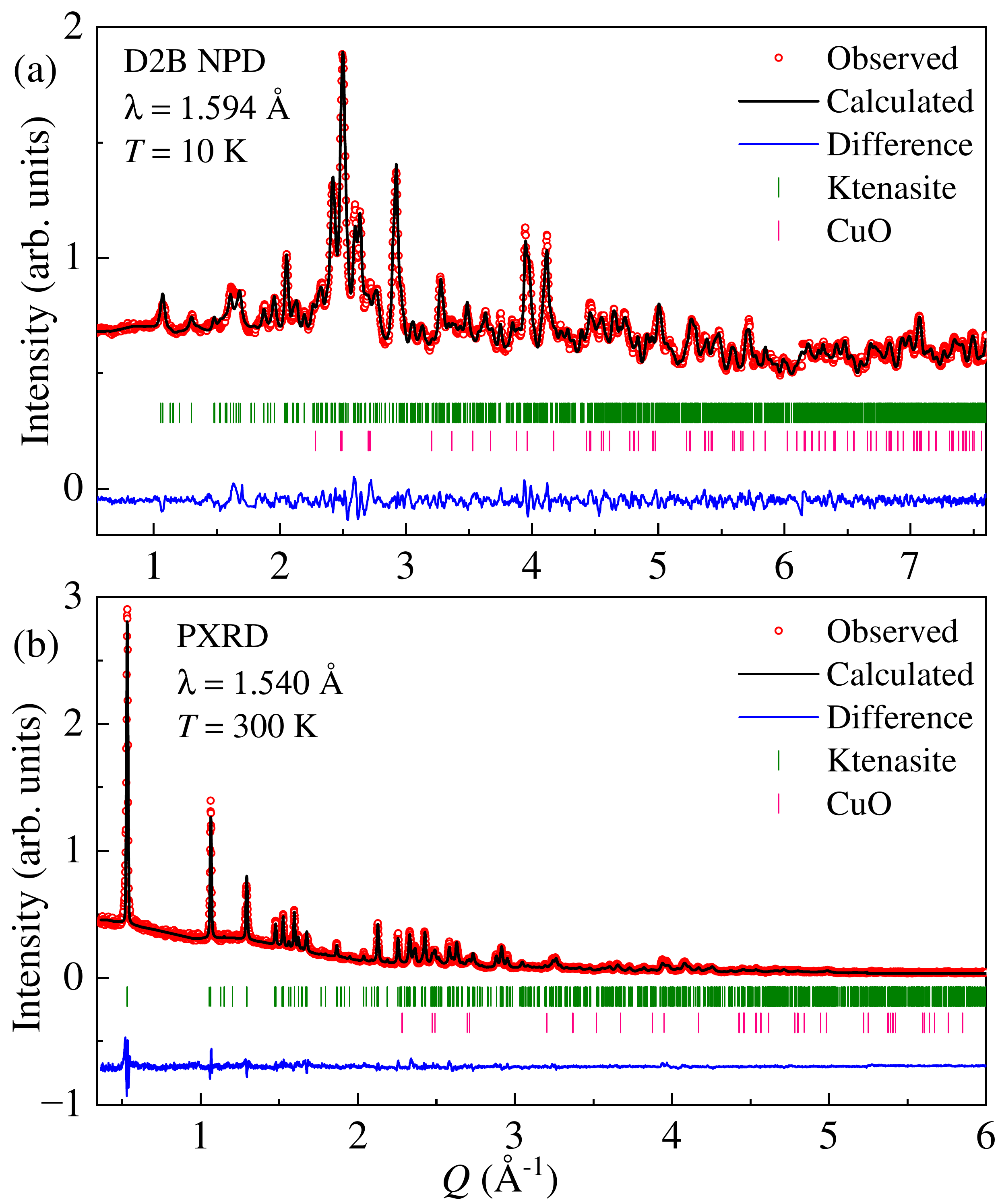}%\vspace{-12pt}
  \caption{\label{XRD_NPD}Rietveld-refined powder (a) neutron  and (b) x-ray patterns for ktenasite.}
\end{figure}

\begin{figure}
  \includegraphics[width=\columnwidth]{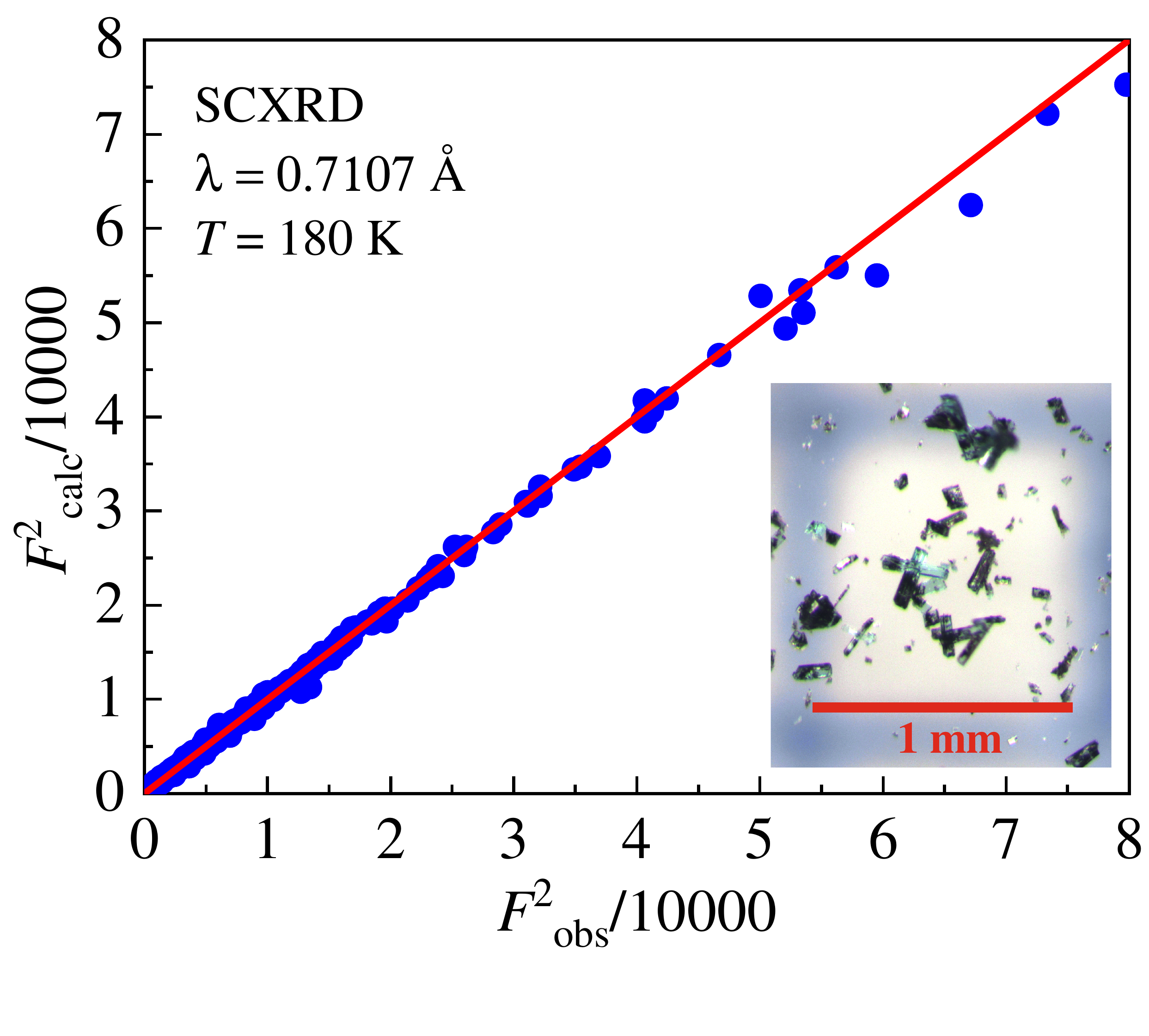}\vspace{-27pt}
  \caption{\label{SCXRD}Structural model fit of single-crystal x-ray diffraction data collected at 180\,K. $F_{\text{calc}}^{2}$ and $F_{\text{obs}}^{2}$
  represent the calculated and observed structure factors, respectively. The inset shows an optical microscope image of several submillimeter-sized ktenasite crystals.}
\end{figure}

The crystal structure of synthetic ktenasite was determined using neutron powder diffraction, x-ray powder diffraction, and single-crystal x-ray diffraction. The Rietveld-refined crystal structure from D2B neutron data at 10\,K is shown in Fig.~\ref{Crystal_Structure}(a). Refinement confirms that the compound crystallizes in monoclinic symmetry with space group $P2_1/c$ (no.\,14), consistent with earlier single-crystal x-ray studies on the natural mineral form~\cite{Mellini1978}. However, our refinement benefits from neutron diffraction data and low-temperature measurements, allowing for more accurate determination of proton positions and subtle structural details not accessible in the 1978 dataset. The Rietveld fits to the powder neutron and x-ray diffraction patterns are presented in Fig.~\ref{XRD_NPD}. In addition to the dominant ktenasite phase, the refinement also reveals a minor impurity phase (2.42\,wt.\%) of unreacted CuO. Given its small quantity and that it is a commensurate antiferromagnet below 213\,K\,\cite{Forsyth1988}, this impurity is unlikely to significantly affect our determination of the intrinsic magnetic properties of ktenasite at low temperature. The crystal structure refinement based on XtremeD data at 11\,K is provided in the Supplemental Material, see Appendix~\ref{supp}. The single-crystal refinement result at 180\,K is shown in Fig.~\ref{SCXRD}, with the inset displaying several sub-millimeter-sized crystals. The refined structural parameters and goodness-of-fit indicators are summarized in Appendix~\ref{appA}, and the corresponding Crystallographic Information Files (CIFs) are available in the ancillary files online.

Ktenasite is composed of corrugated 2D $[\ce{Cu(Cu,Zn)(OH)3O}]^{2-}$ layers built from Jahn–Teller--distorted, edge-sharing octahedra, where each \ce{Cu^{2+}}/\ce{Zn^{2+}} ion is octahedrally coordinated by six oxygen atoms, including hydroxide \ce{OH^-} and oxide \ce{O^{2-}} ligands. On average this corresponds to three \ce{OH^-} groups and one \ce{O^{2-}} ion per metal center per formula unit, accounting for ligand sharing in the extended structure. These layers, illustrated in Fig.\ref{Crystal_Structure}(b), host a 2D triangular arrangement of magnetic \ce{Cu^{2+}} ions. The in-plane Cu–-Cu distances vary between 3.062 and 3.392\,\AA, leading to scalene-type nearest-neighbor interactions, as depicted in Fig.\ref{Crystal_Structure}(c). Connectivity between these layers is achieved through sulfate (\ce{SO4^2-}) tetrahedra which link the octahedral sheets to form composite tetrahedral–octahedral motifs. These stack along the $c$ axis and are interconnected by interlayer \ce{[Zn(H2O)6]^2+} octahedra through hydrogen bonding.

Although previous structural studies on  natural mineral ktenasite samples primarily relied on x-ray diffraction~\cite{Mellini1978}, this method cannot reliably distinguish Cu and Zn due to their similar atomic numbers ($Z$). As a result, the reported mixed occupancies at the Cu1 and Cu2 sites may be inaccurate. In contrast, neutron diffraction offers enhanced element specificity owing to the difference in coherent scattering lengths between Cu ($b_{\text{coh}} = 7.718$\,fm) and Zn ($b_{\text{coh}} = 5.680$\,fm). Using neutron diffraction, we refined site occupancies and found that the Cu1 site is exclusively occupied by Cu, while the Zn layer remains fully Zn, with no detectable site mixing. However, the Cu2 site shows considerable site mixing, with $63.7(5)$\% Zn occupancy.  

\begin{figure*}
  \includegraphics[height=0.45\textheight]{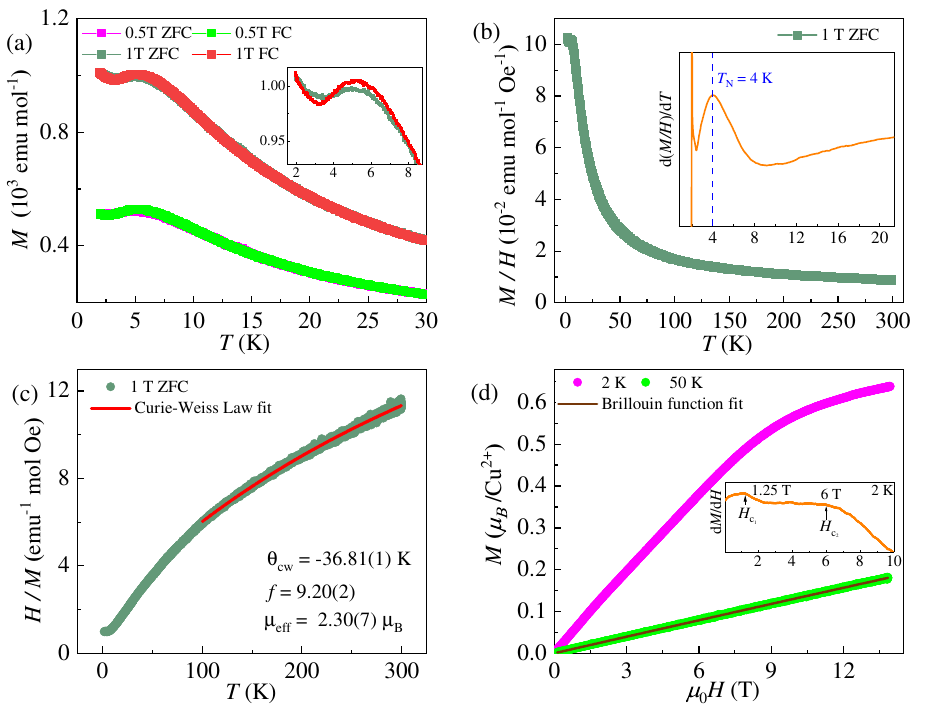}%\vspace{-1.5em}
  \caption{\label{Magnetization}(a) Temperature dependence of \ktH\ magnetization under ZFC and FC conditions at selected fields. The inset is a zoomed-in view showing the onset of ZFC–FC divergence. (b) ZFC $M/H$ as a function of temperature at 1\,T. The inset presents the temperature derivative, revealing a peak near 4\,K. (c) Inverse susceptibility ($H/M$) at $\mu_0H = 1$\,T, with the Curie–Weiss fit shown as a solid red line. (d) Isothermal magnetization at 2 and 50\,K, with Brillouin function fit shown as solid line. The inset shows the field derivative, exhibiting slope changes at 1.25 and 6\,T.}%\vspace{-1.5em}
\end{figure*}

We note that attempts to reduce the Zn content of the samples failed to produce ktenasite.  We suspect that either the presence of Zn on the Cu2 site is essential for stabilizing the structure, or the concentration of Zn required to produce the [Zn(H$_2$O)$_6$]$^{2+}$ layer inevitably leads to considerable substitution on the Cu2 site.  

The magnetic sublattice is influenced by random site disorder, which effectively reduces its dimensionality. In the ideal case, magnetic \ce{Cu^2+} ions form a 2D scalene-distorted anisotropic triangular lattice, as illustrated in Fig.\ref{Crystal_Structure}(c). However, the substantial Zn substitution at the Cu2 site promotes the formation of corrugated 1D spin chains composed solely of \ce{Cu^{2+}} ions at the Cu1 site, as shown in Fig.\ref{Crystal_Structure}(d). The dominant first-neighbor interaction (3.099\,\AA) occurs along the $b$ axis within these chains. Additionally, weak second-neighbor interchain interactions (5.583\,\AA) are possible via magnetic superexchange pathways. With roughly 1/3 occupancy of Cu2 sites by Cu, these links are sporadic and presumably disordered, introducing considerable randomness into the interchain interactions.  Considering the large interlayer separation (11.90\,\AA) and exchange pathways that traverse multiple hydrogen bonds, magnetic interactions along the $c$ axis are expected to be negligible.

To gain deeper insights into the magnetic exchange interactions, we analyze the intra- and interchain superexchange pathways within the framework of the Goodenough-Kanamori-Anderson (GKA) rules~\cite{Goodenough1955, Kanamori1959}. According to these rules, the superexchange between Cu$^{2+}$ ions favors antiferromagnetic (AFM) coupling for Cu--O--Cu bond angles in the range $\sim 95^{\circ}$ to $180^{\circ}$, while bond angles $\lesssim 95^{\circ}$ generally lead to weak or ferromagnetic (FM) interactions.

\begin{table}[b]
  \caption{\label{ExchangePaths}
  Summary of the Cu--O--Cu super-exchange pathways that mediate magnetic interactions both along and between the quasi-1D Cu$^{2+}$ chains, as depicted in Figs.~\ref{Crystal_Structure}(c) and \ref{Crystal_Structure}(d).}
  \begin{ruledtabular}
     \begin{tabular}{lcc}
      Bond & Cu--Cu distance(s) & Cu--O--Cu angle \\ 
       & (\AA) & ($^\circ$) \\ \hline
      \multicolumn{3}{c}{Intrachain ($J_1$)} \\ \hline
      Cu1--O3H2--Cu1 & 3.099 & 97.154 \\
      Cu1--O4H3--Cu1 & 3.099 & 98.529 \\ \hline

      \multicolumn{3}{c}{Interchain ($J'$)} \\ \hline
      Cu1--O1--Cu2        & 3.208, 3.392 & 87.668 \\
      Cu1--O2H1--Cu2      & 3.062, 3.107 & 100.776 \\
      Cu1--O3H2--Cu2      & 3.062, 3.208 & 111.493 \\
      Cu1--O4H3--Cu2      & 3.107, 3.392 & 92.279 \\
    \end{tabular}
  \end{ruledtabular} 
\end{table}

Based on the distances and angle distributions for the dominant paths listed in Table~\ref{ExchangePaths}, the intrachain interactions ($J_1$) are likely to be predominantly AFM. In contrast, the interchain interactions ($J'$) involve both oxide (O$^{2-}$) and hydroxide (OH$^{-}$) bridges. It is well established that superexchange mediated by OH$^{-}$ (or OD$^{-}$) groups is less efficient than via O$^{2-}$ due to reduced orbital overlap and hydrogen-bond-induced distortions. The interchain Cu--O--Cu bond-angles span a wide range (87.7$^{\circ}$ to 111.5$^{\circ}$). Here, the larger angles are expected to mediate moderate AFM couplings, whereas the oxide-bridged path at 87.7$^{\circ}$ is predicted to promote stronger FM interactions. Consequently, the net interchain Cu1--Cu2 coupling is likely FM in nature. Furthermore, several interchain Cu--Cu distances are comparable to, or even shorter than, the intrachain separation, indicating that interchain couplings cannot be entirely neglected. 

Taken together, the structural and GKA analyses point to a quasi-1D magnetic system with residual frustrated interchain interactions, consistent with a coexistence of 1D and weakly 2D magnetic substructures. This scenario naturally explains the observed $T^{2}$ contribution to the low-temperature specific heat in terms of 2D spin-wave excitations (see Sec.~\ref{sec:SpecificHeat}). A quantitative determination of the $J_1/J'$ ratio, however, will require first-principles calculations such as density functional theory.  

Further characterization of the surface morphology, elemental composition, thermal analysis, and bond characteristics are provided in the Supplemental Material in Appendix~\ref{supp}.

\section{{\MakeLowercase{dc}} Magnetization}

Magnetization measurements as a function of temperature and magnetic field were carried out to probe the magnetic interactions in \ktH. Figure~\ref{Magnetization}(a) shows the temperature dependence of the zero-field-cooled (ZFC) and field-cooled (FC) dc magnetization ($M$) at selected magnetic fields ($H$). The inset highlights the thermal hysteresis, evident from the ZFC–FC divergence at 1\,T, indicative of frozen spins in the system. Figure~\ref{Magnetization}(b) presents the ZFC $M/H$ versus temperature at 1\,T, which increases smoothly upon cooling and displays a broad hump around 5.5\,K. The temperature derivative of $M/H$, shown in the inset, exhibits a peak near 4\,K, suggesting the possible onset of LRO coexisting with a spin-glass state. The high-temperature susceptibility data (100–300\,K) under 1-T ZFC conditions were fit to the Curie–Weiss law given by
\setlength{\abovedisplayskip}{5pt}
\setlength{\belowdisplayskip}{5pt}
\begin{align}
\chi = \chi_0 + \frac{C}{(T - \theta_{\text{CW}})},
\end{align}
where $\chi_0$ accounts for temperature-independent contributions such as diamagnetism and van~Vleck paramagnetism, and $C$ and $\theta_{\text{CW}}$ are the Curie constant and Weiss temperature, respectively. As shown in Fig.~\ref{Magnetization}(c), the data follow Curie-Weiss behavior, yielding $\chi_0 = 3.483\times10^{-3}$\,emu/mol$_\text{Cu}$, $\theta_{\text{CW}} = -36.81(1)$\,K, and $C = 0.66(5)$\,emu\,K/mol$_\text{Cu}$\,Oe. No sign of the magnetic transitions of the CuO impurity at 213 and 230\,K is observed. The negative $\theta_{\text{CW}}$ indicates predominant AFM interactions between $\text{Cu}^{2+}$ spins. The effective magnetic moment is $\mu_{\text{eff}} = 2.30(7)\,\mu_{\text{B}}/\text{Cu}^{2+}$, exceeding the spin-only $S=1/2$ value of $1.73\,\mu_{\text{B}}/\text{Cu}^{2+}$, likely reflecting distortion-enhanced unquenched orbital contributions. The frustration factor, defined as $f = \lvert \theta_{\mathrm{CW}} \rvert / T_{\mathrm{N}}$, quantifying the degree of magnetic frustration is $\sim$\,9.20(2) highlighting strong frustration. 

Figure~\ref{Magnetization}(d) shows the isothermal magnetization curves at 2 and 50\,K. For the 50\,K data, the data remain linear throughout the measured field range; on the other hand for the 2\,K data, below 8\,T, $M$ increases approximately linearly, while above that, it begins to saturate, indicating weak exchange interactions or low connectivity in the magnetic sublattice in \ktH. The magnetization reaches a maximum of  0.64\,$\mu_{\text{B}}/$\text{Cu}$^{2+}$ at 2\,K, which constitutes only 64$\%$ of the expected saturation magnetization \mbox{$M_{\text{s}} = g_JJ\mu_{\text{B}}= 1\,\mu_{\text{B}}/$\text{Cu}$^{2+}$} for $\text{Cu}^{2+}$ ions. A CuO impurity level of 2.42\,wt.\% corresponds to at most $\sim$\,8\% of the Cu spins and would contribute no more than $\sim$\,0.08\,$\mu_{\text{B}}/\text{Cu}^{2+}$ to $M(H)$, which is negligible. The incomplete polarization at 14\,T may stem from the presence of frozen correlated spins. The isothermal magnetization curve in the paramagnetic state at 50\,K is fitted using the relation ${M} = {M_s}B_J(y)$, where the Brillouin function $B_J(y)$ is
\setlength{\abovedisplayskip}{5pt}
\setlength{\belowdisplayskip}{5pt}
\begin{align}
B_J(y) = \left[\frac{2J+1}{2J}\coth\left(\frac{y(2J+1)}{2J}\right) - \frac{1}{2J}\coth\left(\frac{y}{2J}\right)\right].
\end{align}
Here, $y = {g_J\mu_BJ\mu_0H}/{k_\text{B}T}$, with $g_J$ representing the Land{\'e} $g$ factor. For the fit, $J$ was fixed at $1/2$, leaving $g_J$ as the sole adjustable parameter. The solid line in Fig.~\ref{Magnetization}(d) represents the Brillouin function fit, yielding a $g_J$ factor of 1.98, close to the theoretical value of 2. The 2-K data are poorly described by the Brillouin function, as expected, since it does not account for magnetic correlations and spin-freezing effects. No saturation was seen up to 14\,T at 2\,K, with the field derivative in the inset indicating slope changes suggestive of field-induced transitions around 1.25 and 6\,T.

\section{\MakeLowercase{ac} Susceptibility}

\begin{figure*}[t]
  \includegraphics[width=\textwidth]{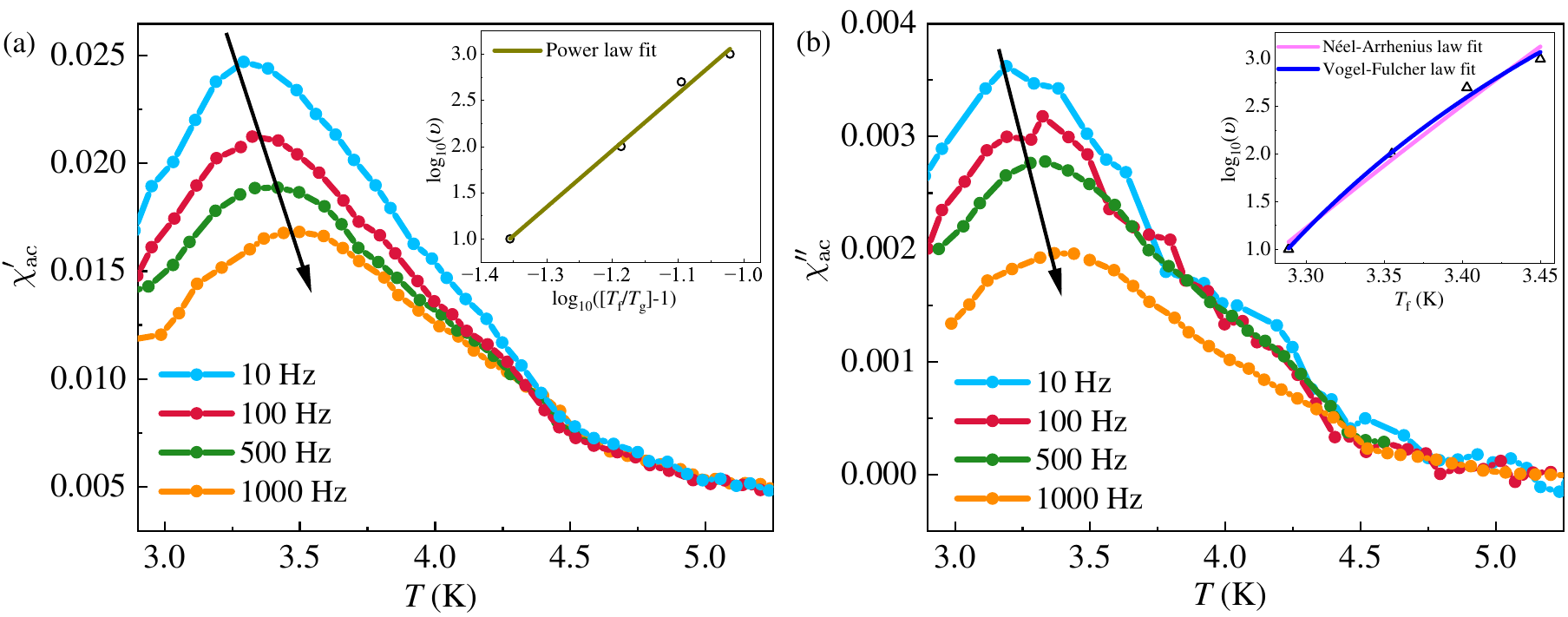}%\vspace{-1.5em}
  \caption{\label{acSusceptibility}Temperature and frequency dependence of the (a) real ($\chi^{\prime}_{\mathrm{ac}}$) and (b) imaginary ($\chi^{\prime\prime}_{\mathrm{ac}}$) components of the ac susceptibility. The inset in (a) shows the frequency dependence of the freezing temperatures fitted using a power law, while the inset in (b) shows fits using both N\'eel-Arrhenius and Vogel-Fulcher laws.}%\vspace{-1.5em}
\end{figure*}

ac Susceptibility measurements were conducted to investigate potential spin freezing in this structurally disordered material, as suggested by the bifurcation of ZFC and FC dc magnetization. Figures~\ref{acSusceptibility}(a) and \ref{acSusceptibility}(b) show the temperature dependence of the real ($\chi^{\prime}_{\mathrm{ac}}$) and imaginary ($\chi^{\prime\prime}_{\mathrm{ac}}$) parts of the ac susceptibility measured at various frequencies. Both components exhibit a frequency-dependent shift of the broad peak (centered near 3.28\,K at frequency $\nu$ = 10\,Hz) towards higher temperature at higher frequency, indicative of glassy dynamics~\cite{Binder1986, Mydosh1993}. The relative dissipation ($\chi^{\prime\prime}_{\mathrm{ac}}/\chi^{\prime}_{\mathrm{ac}}$ $\sim$\,0.1) aligns with expectations for insulating spin glasses. A possible weak jump is seen in both components around 4.3\,K, above which both components become frequency independent. The position of the peak in ($\chi^{\prime}_{\mathrm{ac}}$) is taken as the freezing temperature ($T_\mathrm{f}$), below which a crossover from dynamic to quasi-static spin behavior occurs. In addition to the shift to higher temperatures, there is peak intensity reduction with increasing frequency, consistent with spin-relaxation effects arising due to a broad distribution of relaxation times in spin glasses.

\begin{figure*}
  \includegraphics[height=0.45\textheight]{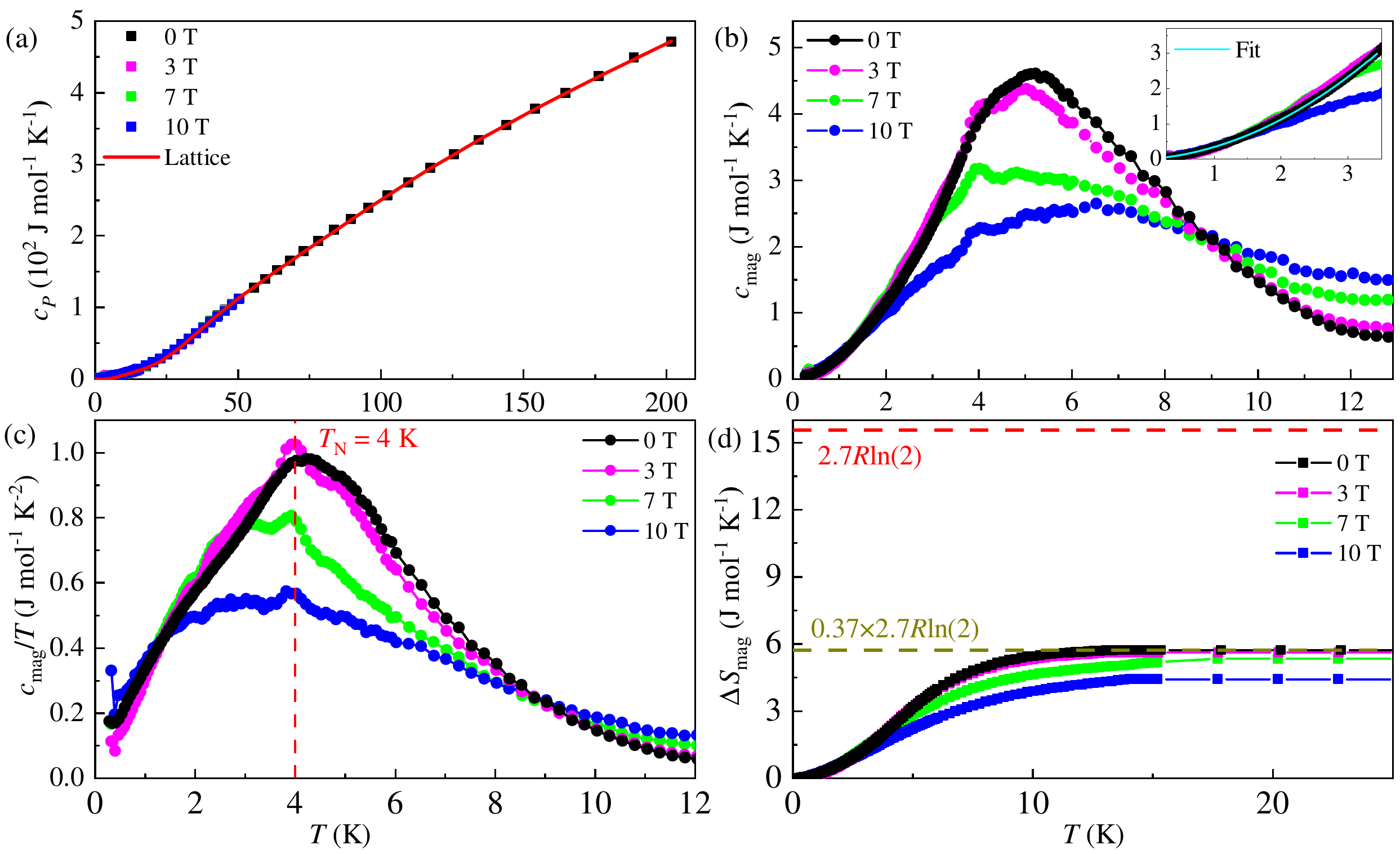}%\vspace{-1.5em}
  \caption{\label{Specific_heat}(a) Temperature dependence of the total specific heat (\Cp) of \ktH, measured down to 350\,mK under various magnetic fields. The solid line shows the lattice contribution, fitted using the Debye–Einstein model in the 15–200\,K temperature range. (b) Magnetic specific heat (\Cmag) as a function of temperature for different magnetic fields, with the solid line in the inset representing the fit based on Eq.~\ref{cmag}. (c) Temperature dependence of $\Cmag/T$ and (d) magnetic entropy (\Smag) under various magnetic fields.}%\vspace{-1.5em}
\end{figure*}

The Mydosh parameter ($\phi$) quantitatively characterizes spin-glass behavior based on the relative shift in the freezing temperature with frequency~\cite{Mulder1981, Mydosh1993}. It is defined as:
\begin{equation}
\phi = \frac{\Delta T_\mathrm{f}}{T_\mathrm{f} \, \Delta \log_{10} (\nu)},
\end{equation}
where $\Delta \log_{10}(\nu) = \log_{10}(\nu_2) - \log_{10}(\nu_1)$ and $\Delta T_\mathrm{f} = T_\mathrm{f}({\nu_2}) - T_\mathrm{f}({\nu_1})$. Using $\Delta T_\mathrm{f} = 0.161$\,K and $\Delta \log_{10}(\nu) = 2$, we obtain $\phi = 0.0245$. 
The average value of $\phi_\text{avg} = 0.0266$ falls within the typical range of $\phi\sim$\,0.005--0.05 observed in classical spin-glass and interacting-particle systems. Importantly, this excludes the possibility of superparamagnetic behavior, which typically yields much larger values ($\phi>$~0.1)~\cite{Dormann1988,Mydosh1993}. Depending on the microscopic origin of the magnetic freezing, systems can be classified as canonical spin glasses ($\phi\lesssim$~0.01) where spins freeze cooperatively~\cite{Souletie1985,Mydosh1993,Gondh2021,Banerjee2023}, or cluster spin glasses ($\phi\sim$\,0.02-0.05), involving interacting spin clusters~\cite{Mydosh1993,Anand2012,Stimpson2018,Bag2018,Bhatt2023,Banerjee2023}. In our case, the value of $\phi$ clearly supports a cluster spin-glass nature in \ktH, likely enhanced by competing interactions and structural disorder. 

According to the standard dynamical scaling theory~\cite{Hohenberg1977, Gunnarsson1988}, spin dynamics exhibit critical slowing down, with the relaxation time diverging at the spin-glass transition temperature ($T_\mathrm{g}$) where ergodicity is broken. This behavior is captured by the scaling relation
\begin{equation}
\tau = \tau_0 \left( \frac{T_\mathrm{f}(\nu)}{T_\mathrm{g}}-1 \right)^{-z \nu'},
\label{scaling_relation}
\end{equation}
where $\tau = (2 \pi \nu)^{-1}$ is the relaxation time of the spin fluctuations related to  $\nu$, and $T_\mathrm{f}(\nu)$ is the frequency-dependent freezing temperature. In the low-frequency limit ($\nu \to 0$), $T_\mathrm{f}$ approaches the intrinsic spin-glass transition temperature $T_\mathrm{g}$. Here, $\tau_0$ is the microscopic characteristic time scale, $z$ is the dynamical scaling exponent, and $\nu'$ is the critical exponent which characterizes the divergence of the spin-spin correlation length $\xi$, given by $\xi = \left( \frac{T_\mathrm{f}}{T_\mathrm{g}} - 1 \right)^{-\nu'}$. Since $\tau \propto \xi^z$, this leads to the commonly used linearized power-law form for fitting the frequency dependence of $T_\mathrm{f}$:
\begin{equation}
\log_{10}(\nu) = \log_{10}(\nu_0) + z\nu'\cdot\log_{10} \left( \frac{T_\mathrm{f}}{T_\mathrm{g}}-1 \right).
\label{power_law}
\end{equation}
As shown in the inset of Fig.~\ref{acSusceptibility}(a), the best fit to Eq.~\ref{power_law} yields $T_\mathrm{g}=3.15\,$K, $z\nu'=6.15$ and $\tau_0=4.65 \times 10^{-10}$\,s. These values lie well within the typical range for spin-glass systems: $z\nu$ generally falls between 4 and 12~\cite{Souletie1985, Kumar2021, Gondh2021, Banerjee2023}, while $\tau_0$ spans $10^{-12}$--$10^{-14}$\,s for canonical spin glasses~\cite{Mydosh1993, Gondh2021} and $10^{-7}$--$10^{-10}$\,s for cluster spin glasses~\cite{Stimpson2018, Bag2018,Kumar2021, Bhatt2023}.

The extracted parameters thus confirm that \ktH\,~exhibits cluster spin-glass behavior, with slow spin dynamics arising from interactions among spin clusters rather than isolated magnetic moments. The relatively large $\tau_0$ further supports this scenario, indicating the presence of mesoscopic, correlated dynamics driven by intercluster interactions.

To further verify the nature of spin relaxation, we attempted to apply the N\'eel-Arrhenius law, which is typically valid for non-interacting or weakly interacting magnetic moments, given by~\cite{Binder1986}
\begin{equation}
\tau = \tau^*\exp\left(\frac{E_\mathrm{a}}{k_\mathrm{B} T_\mathrm{f}}\right),
\label{neel_arrhenius}
\end{equation}
where $\tau^*$ represents the characteristic spin-flip time (analogous to $\tau_0$), and $E_\mathrm{a}$ is the average activation energy barrier. The activation energy quantifies the energetic separation between metastable spin configurations, and the Arrhenius law models the thermally activated nature of such transitions.
As shown in the inset of Fig.~\ref{acSusceptibility}(b), although the fit to Eq.~\ref{neel_arrhenius} yields a reasonably good linear fit, the extracted parameters $\tau^*=7.59\times10^{-46}$\,s and $E_\mathrm{a}/k_\mathrm{B}=334(34)$\,K are unphysical. This failure of the  N\'eel-Arrhenius model further confirms that the observed spin relaxation does not arise from independent or weakly interacting spins, but must instead result from cooperative freezing due to intercluster interactions, consistent with cluster spin-glass behavior.

With the presence of intercluster interactions established, we apply the Vogel-Fulcher law, which incorporates such interactions and is expressed as~\cite{Souletie1985}:
\begin{equation}
\tau = \tau_0 \exp\left(\frac{E_a}{k_\text{B} (T_\mathrm{f} - T_0)}\right),
\label{vogel_fulcher}
\end{equation}
where $T_0$ quantifies the strength of the intercluster interactions. As shown in the inset of Fig.~\ref{acSusceptibility}(b), the Vogel-Fulcher--law fit yields good agreement with the data. Using the previously obtained value of $\tau_0$ from the dynamical scaling analysis (Eq.~\ref{power_law}), we extract $E_\mathrm{a}/k_\mathrm{B}=7.37$\,K and $T_0=2.86$\,K. Unlike the N\'eel–Arrhenius fit, these values are physically reasonable, reinforcing the importance of finite intercluster coupling. The ratio $E_\mathrm{a}/k_\mathrm{B}T_0 = 2.58$ lies in the intermediate regime, indicating moderate interaction strength among magnetic clusters. This further supports the interpretation that the observed spin relaxation arises from cooperative freezing in a system of correlated magnetic clusters.

\section{Specific Heat}\label{sec:SpecificHeat}

Specific heat (\Cp) measurements were performed over a wide range of temperatures and magnetic fields to probe potential magnetic transitions, low-energy excitations and field-dependent thermodynamic behavior in \ktH. The temperature dependence of \Cp\ under various fields is shown in Fig.~\ref{Specific_heat}(a). Attempts to synthesize a  nonmagnetic, isostructural Zn-based analog were unsuccessful, yielding a phase mixture of lahnsteinite [Zn$_4$(SO$_4$)(OH)$_6\cdot$3H$_2$O] and namuwite [Zn$_4$(SO$_4$)(OH)$_6\cdot$4H$_2$O]. In the absence of a suitable reference sample, the magnetic contribution to the specific heat (\Cmag) was estimated by subtracting a modeled lattice contribution from the total \Cp. The lattice contribution was approximated fitting a Debye-Einstein model over the 15–200\,K range, where phonons dominate. This model comprises one Debye and four Einstein components, expressed as
\begin{align}
c_{\text{lattice}}(T) &= f_\text{D} c_\text{D}(\theta_\text{D}, T) + \sum_{i=1}^{4} g_i c_{\text{E}_i}(\theta_{\text{E}_i}, T).
\label{clattice}
\end{align}
The first term in Eq.~\eqref{clattice} represents the Debye model, accounting for acoustic modes:
\begin{align}
c_\text{D}(\theta_\text{D}, T)~=~&9NR \left( \frac{T}{\theta_\text{D}} \right)^{\!3}\!\int_0^{\theta_\text{D}/T}\!\frac{x^4 e^x}{(e^x - 1)^2}\,{\rm d}x. \notag \\
\end{align}
Here, $\theta_\text{D}$ represents the Debye temperature, $N$ is the total number of atoms in the formula unit, $R$ is the universal gas constant, and $x$ is defined as $\frac{\hbar \omega}{k_\text{B} T}$. The second, Einstein, term in Eq.~\eqref{clattice} accounts for the optical modes:
\begin{align}
c_\text{E}(\theta_\text{E}, T)~=~&3NR \left( \frac{\theta_\text{E}}{T} \right)^2 \frac{e^{\theta_\text{E}/T}}{(e^{\theta_\text{E}/T} - 1)^2},
\end{align}
where $\theta_\text{E}$ is the Einstein temperature. Since the formula unit contains 45 atoms, each contributing three vibrational degrees of freedom, the total number of phonon modes is 135. The three acoustic modes were represented by the Debye term, fixing the weight factor to $f_{\text{D}}=3/135\simeq0.022$. The remaining 132 optical modes were distributed among the four Einstein terms, with $\sum_{i} g_i = 132/135$, ensuring consistency with the Dulong–Petit limit at high temperatures. As seen in Fig.~\ref{Specific_heat}(a), the resulting fit reproduces the experimental data quite well, yielding $\theta_\text{D}=65$\,K, $g_1=0.312$, $g_2=0.305$, $g_3=0.177$, $g_4=0.183$ $\theta_{\text{E}_1}=1230$\,K, $\theta_{\text{E}_2}=481$\,K, $\theta_{\text{E}_3}=163$\,K and $\theta_{\text{E}_4}=1337$\,K.

 The resulting \Cmag, shown in Fig.~\ref{Specific_heat}(b), exhibits a  broad maximum ($T_\mathrm{peak}$) around 5.36\,K, likely due to spin freezing and/or short-range order, accompanied by a sharp $\lambda$-type anomaly at 4\,K that signals the onset of LRO in zero field. The CuO contribution to the specific heat is limited to a few percent by mass and cannot account for the observed 4\,K anomaly. Notably, the broad peak occurs at a higher temperature than the spin freezing point, with $T_\mathrm{peak} \approx 1.6 T_\mathrm{f}$\,---\,a characteristic feature of spin-glass systems~\cite{Binder1986}. Upon increasing the magnetic field, the broad feature is progressively suppressed and shifts to lower temperatures, eventually vanishing at 10\,T, indicating that the field destabilizes the spin-glass state. Below \TN, the magnetic specific heat can be described by the empirical form
\begin{align}
c_\text{mag}(T) = \alpha T + \beta T^2,
\label{cmag}
\end{align}
 as shown in the inset of Fig.~\ref{Specific_heat}(b), with $\alpha = 0.134$\,J\,/(mol K$^2$) and $\beta = 0.214$\,J\,/(mol K$^3$) in zero field. The linear term reflects the spin-glass contribution, arising from localized spin excitations and tunneling between metastable states, which yield an energy-independent flat density of states. The quadratic term corresponds to gapless spin-wave excitations, as expected for Goldstone modes with linear dispersion ($\omega \propto |k|$) in a quasi-2D antiferromagnet with continuous spin-rotational symmetry. As shown in Fig.~\ref{Specific_heat}(c), the $\lambda$-like anomaly in $\Cmag/T$ remains sharp and essentially field-independent, confirming the robustness of the long-range ordered state.  This sharp peak may correspond to the weak jump in the ac susceptometry just above 4\,K in Figs.~\ref{acSusceptibility}(a) and \ref{acSusceptibility}(b).

The magnetic entropy (\mbox{$\Smag = \int_{T'}^T \frac{\Cmag}{T} \text{d}T$},  where $T'$ is the lowest measured temperature) in Fig.~\ref{Specific_heat}(d) saturates around $\sim$\,20~K, reaching approximately 37\% of the expected value $2.7R\ln(2) = 15.56\,\text{J/mol}\cdot\text{K}$ for \(S = \frac{1}{2}\) spins in \ktH. Such a substantial entropy reduction is typical in frustrated systems \cite{Mannathanath_Chakkingal2025,Parui2025,Kulbakov2025b,Kumar2025, Akshay2025}, and may stem from lattice model limitations, disorder, spin-glass behavior, or unaccounted-for contributions at ultralow temperatures.

\section{Magnetic diffraction}

\begin{figure}
  \includegraphics[width=\columnwidth]{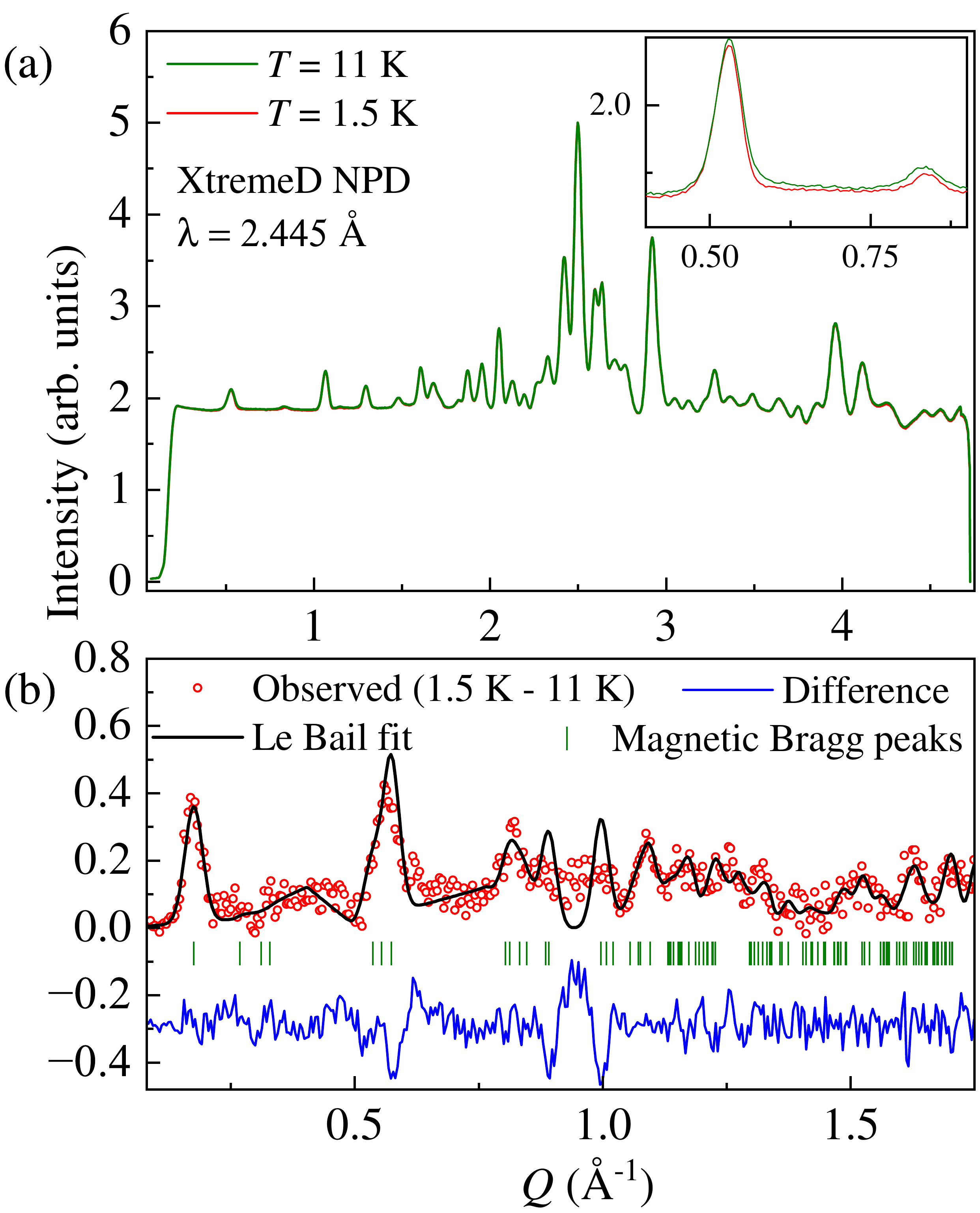}%\vspace{-1.5em}
  \caption{\label{Mag_diffraction}(a) Neutron powder diffraction patterns collected at 11 and 1.5\,K using XtremeD. The inset provides an enlarged view of the low-$Q$\ region showing magnetic features that emerge at low temperatures. (b) Difference pattern highlighting sharp magnetic Bragg peaks emerging in the low-$Q$\ region together with the corresponding Le Bail fit.}%\vspace{-1.5em}
\end{figure}

\begin{figure*}
\includegraphics[width=0.85\textwidth]{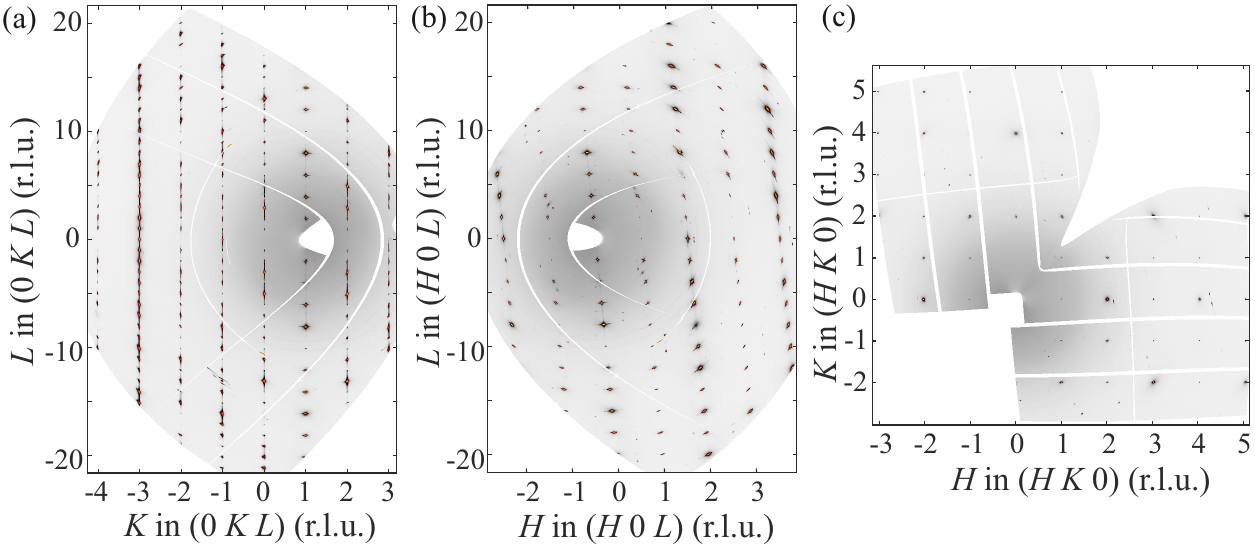}
\caption{\label{diffuse}Cuts through the x-ray diffuse scattering in synthetic ktenasite:  the (a) $(0\,K\,L)$, (b) $(H\,0\,L)$, and (c) $(H\,K\,0)$ planes.}
\end{figure*}

To conclusively determine the presence of magnetic LRO coexisting with the cluster spin-glass state, we conducted low-temperature neutron diffraction measurements to search for sharp magnetic Bragg peaks indicative of spin ordering. Figure~\ref{Mag_diffraction}(a) shows the neutron diffraction patterns collected above and below \TN. An inset in Fig.\ref{Mag_diffraction}(a) provides a magnified view of the low-$Q$ region, highlighting magnetic features that become prominent at low temperatures. The difference pattern in Fig.~\ref{Mag_diffraction}(b) further emphasizes these features, revealing sharp magnetic Bragg peaks in the low-$Q$\ region, consistent with the development of LRO. Notably, the magnetic peak intensity decreases rapidly with increasing $Q$, in agreement with the \ce{Cu^{2+}} magnetic form factor. These findings confirm that our synthetic ktenasite indeed exhibits LRO coexisting with a cluster spin-glass state, as independently supported by dc magnetization, specific heat, and ac susceptibility measurements. 

To gain deeper insight into the magnetic structure, we employed the $k$-search program from the \textsc{FullProf} Suite~\cite{FullProf} to determine the magnetic propagation vector, obtaining $\mathbf{k} = \left(0.015\ 0.1605\ 0.00045\right)$ with an R-factor of 1.27\,\% based on the four most intense magnetic Bragg peaks. This vector was then used as the starting point for the Le Bail fit [Fig.~\ref{Mag_diffraction}(b)], which reproduced the data well and yielded a refined propagation vector of $\mathbf{k} = \left(0.000(3)\ 0.1696(3)\ 0.040(5)\right)$. The exact nature of the incommensurate order cannot be uniquely determined from the present data and will be addressed in a dedicated follow-up study. A symmetry-based consistency check of the incommensurate magnetic order is provided in the Supplemental Material, see Appendix~\ref{supp}.

\section{Diffuse Scattering}

One explanation for the coexistence of a sharp specific heat transition with a broad glassy transition would be local Zn/Cu order, whereby a cation-ordered phase would have a sharp transition to long-range order while the disordered parts of the sample would contribute the glassy behavior.  To examine this possibility, we performed x-ray diffuse scattering on a small single crystal at the ID28 beamline at the ESRF in Grenoble, France.  

High-symmetry cuts through the diffuse scattering are shown in Fig.~\ref{diffuse}.  We observe no hints of diffuse scattering which could suggest a tendency toward local cation order.  The relatively low x-ray contrast between Cu and Zn does not allow us to definitively exclude local order based on this data, but any short-range cation order would need to be extremely weak and have essentially no effect on the oxygen ligands.  

\section{Discussion}

The magnetism in ktenasite emerges from a complex interplay of competing interactions, geometric frustration, and site disorder. As shown in Fig.~\ref{Crystal_Structure}(b) and (c), the Cu1 and Cu2 sites comprise an anisotropic 2D triangular lattice, giving rise to frustrated 2D magnetism. However, occupation of the Cu2 site mainly by non-magnetic Zn leaves the Cu1 sublattice to form 1D spin chains. Neutron diffraction confirms site disorder at the Cu2 site, with partial Cu/Zn occupancy, and diffuse x-ray scattering failed to find signs of local cation order, indicating that the system is a statistical mixture of 2D and 1D magnetic units.

The emergence of a glassy magnetic state is therefore unsurprising, as the system hosts both essential ingredients:
(i) Disorder, from random Cu/Zn occupation, which renders the formation of triangular or chain motifs probabilistic and introduces randomized exchange interactions; and
(ii) Frustration, from the triangular geometry, resulting in highly degenerate spin configurations.

Interestingly, despite this disorder, LRO still emerges. This could be due to strong exchange coupling between Cu$^{2+}$ ions at the Cu1 sites, which may be robust enough that disorder alone cannot fully disrupt magnetic correlations. Alternatively, it is also possible that the Cu$^{2+}$ network remains above the percolation threshold, maintaining sufficient magnetic connectivity through both Cu1 and partially occupied Cu2 sites. Theoretical studies on random-site percolation in triangular lattices \cite{Malarz2021} demonstrate that, even with complex local environments, magnetic connectivity can persist above a critical threshold determined by coordination. A comparable percolation-driven magnetic ordering was recently reported in Cu$_4$(OH)$_6$Cl$_2$~\cite{Zheng2024}, where static spin clusters percolated to form a coherent long-range-ordered state. In ktenasite, the reduced-dimensional 1D spin chains and the percolating 2D triangular motifs (formed when the Cu2 site is occupied by Cu), likely help stabilize the long-range correlations. Thus, the system supports LRO in regions where more magnetic pathways are intact, while other disordered regions, disrupted by Zn substitution, give rise to glassy spin clusters consistent with the ac susceptibility analysis.

It is noteworthy that this long-range ordered state is remarkably robust: as shown in Fig.\ref{Specific_heat}(c), the $\lambda$-like anomaly remains sharp and insensitive to applied magnetic fields up to 10\,T. This robustness contrasts sharply with the behavior of structurally disordered quantum systems such as the ludwigites \ce{Cu2$M$^{$\prime$}BO5} ($M^{\prime}$ = Al, Ga)\cite{Kulbakov2021}, where a structurally ordered \ce{Cu^{2+}} ladder sublattice is interpenetrated by a randomly occupied sublattice of magnetic \ce{Cu^{2+}} and nonmagnetic \ce{Ga^{3+}}/\ce{Al^{3+}} ions. In those systems, even a small magnetic field of 1\,T completely suppresses LRO, driving a crossover into a spin-glass regime. In ktenasite, however, the dominant magnetic interactions among spin chains and triangular motifs appear to outweigh the Zeeman energy within the explored field range, thereby mitigating the destabilizing effects of weak disorder or orphan spins. This balance enables not only the persistence of LRO but also its coexistence with glassy magnetic features.

\section{Conclusion}

In summary, ktenasite [\ktH] exhibits a rare dual magnetic ground state, where a cluster spin-glass coexists with incommensurate LRO. Structural characterization shows significant Cu/Zn site disorder, which drives a dimensional crossover from a 2D scalene-distorted triangular lattice to intertwined 1D spin chains. Magnetization, ac susceptibility, and specific heat measurements confirm glassy behavior alongside sharp anomalies at 4\,K from LRO, while neutron diffraction reveals well-defined magnetic Bragg peaks. Remarkably, unlike structurally disordered ludwigites, the LRO in ktenasite remains robust up to 10\,T, demonstrating that disorder can stabilize rather than merely suppress magnetic order. This coexistence, arising from percolation within a spatially intertwined network of 2D and 1D magnetic substructures, establishes ktenasite as a rare experimental platform linking frustration, spin-glass physics and dimensional crossover, highlighting how disorder can tune and stabilize competing quantum phases.

Open questions remain regarding whether increasing disorder or Zn substitution could further disrupt coupling between the 1D spin chains to realize novel quantum magnetic states. While attempts to reduce the Zn concentration were not successful, it may be possible to increase it, or to stabilize the structure with other nonmagnetic $M^{2+}$ ions which will not cross-substitute with Cu, in order to tune the disorder and dimensionality. It would be useful to further investigate the dual ground state of \ktH\ using local probes such as muon spin relaxation or NMR to disentangle the static and dynamic magnetic components. Additionally, theoretical work is needed to understand the magnetism in disorder-free ktenasite. The growth of larger single crystals would facilitate anisotropic measurements and advanced neutron scattering experiments, including inelastic scattering, to determine the magnetic exchanges. 
  
\section*{Data Availability}

Samples and data are available upon reasonable request from D.~C. Peets or D.~S. Inosov; data underpinning this work is available from Refs.~\onlinecite{dataset_ktenasite,Parui2025b,Easy-1475}. Supplementary crystallographic data measured at 180\,K can be retrieved from the joint CCDC/FIZ Karlsruhe structural database (\url{https://www.ccdc.cam.ac.uk/structures/}) by quoting the deposition number CSD-2486479.

\begin{acknowledgments}
We gratefully acknowledge E. Hieckmann for her assistance with the SEM/EDS experiments and I. Kunert for her support with the TG-DTA measurements. This project was funded by the Deutsche Forschungsgemeinschaft (DFG, German Research Foundation) through: individual grants IN 209/12-1, DO 590/11-1 (Project No.\ 536621965), and PE~3318/2-1 (Project No.\ 452541981); through projects B03, C01, and C03 of the Collaborative Research Center SFB~1143 (Project No.\ 247310070); and through the W\"urzburg-Dresden Cluster of Excellence on Complexity and Topology in Quantum Materials\,---\,\textit{ct.qmat} (EXC~2147, Project No.\ 390858490). A.K. acknowledges financial support from the Education Department of the Basque Government under Grant No.\ PIBA-2023-1-0051. The PPMS Dynacool-12 at TUBAF was funded through DFG Project No.\ 422219907. The authors acknowledge the support of the Institut Laue-Langevin, Grenoble, France. 
\end{acknowledgments}

\appendix
\section{Crystal Structure Refinement Details\label{appA}}

Details of our crystal structure refinements are summarized in Table~\ref{Summary}. The refined atomic positions of ktenasite obtained from powder data using neutrons and x-rays are listed in Tables~\ref{NPD_D2B} and \ref{XRD_Stoe}, respectively. Tables~\ref{SCXRD_Ktenasite} and \ref{SCXRD_Ktenasite2} present the crystal data and structure refinement results from SCXRD, together with the refined atomic positions. CIF files describing these refinements are provided in the ancillary files as part of this arXiv submission.

\begin{table}[!htbp]
  \caption{\label{Summary}Summary of crystal structure refinements of \mbox{ktenasite} using PXRD, SCXRD and NPD.}
  \begin{ruledtabular}
    \begin{tabularx}{\columnwidth}{l@{\extracolsep{\fill}}ccc}
      Parameter & PXRD & SCXRD & NPD \\ \hline
      Space group & $P2_1/c$ (no. 14) & $P2_1/c$ (no. 14) & $P2_1/c$ (no. 14) \\
      $T$ (K) & 300 & 180 & 10 \\
      $a$ (\AA) & 5.5945(3) & 5.5882(2) & 5.5825(2) \\
      $b$ (\AA) & 6.1681(1) & 6.1730(2) & 6.1563(6) \\
      $c$ (\AA) & 23.7253(7) & 23.6393(7) & 23.5646(3) \\
      $\beta$ ($^\circ$) & 95.3650(1) & 95.570(3) & 95.5798(3) \\
      $V$ (\AA$^3$) & 815.12(4) & 811.609(9) & 806.036(4) \\
      $Z$ & 2 & 2 & 2 \\
      Density (g\,cm$^{-3}$) & 3.007(1) & 2.968(3) & 3.043(5) \\
      $Q$ range (\AA$^{-1}$) & 0.36--6.00 & 1.05--9.71 & 0.55--7.60 \\
      $R$ (\%) & 4.91 & 3.59 & 2.08 \\
      $wR$ (\%) & 6.58 & 7.37 & 2.73 \\
    \end{tabularx}
  \end{ruledtabular}
\end{table}

\begin{table}[tb]
  \caption{\label{NPD_D2B}Refined atomic positions in \ktD\ at 10\,K from neutron powder diffraction on D2B using 1.594-\AA\ neutrons. The Zn1 site occupies the 2$a$ Wyckoff position, while all other sites occupy 4$e$. The deuteration level refined to 76.80(3)\,\%. }
\begin{tabular}{l r@{.}l r@{.}l r@{.}l c}\hline\hline
Site & \multicolumn{2}{c}{$x$} & \multicolumn{2}{c}{$y$} & \multicolumn{2}{c}{$z$} & Occ. \\ \hline
Zn1   & 0&00000  & 0&00000  & 0&00000  & 1.000 \\
Cu1   & $-$0&02450(2) & 0&11401(4)  & 0&24427(3)  & 1.000 \\
Cu2   & 0&48890(4)  & $-$0&12370(6) & 0&25212(2)  & 0.363(5) \\
Zn2   & 0&48890(4)  & $-$0&12370(6) & 0&25212(2)  & 0.637(5) \\
S1    & 0&37124(8)  & 0&07085(5)  & 0&37571(7)  & 1.000 \\
O1    & 0&36646(2)  & 0&12519(6)  & 0&31049(3)  & 1.000 \\
O2    & 0&61470(4)  & 0&12218(4)  & 0&21236(8)  & 1.000 \\
O3    & 0&81922(1)  & 0&37557(4)  & 0&28511(2)  & 1.000 \\
O4    & 0&14587(3)  & 0&34421(2)  & 0&20860(11) & 1.000 \\
O5    & 0&14831(2)  & 0&02792(8)  & 0&39958(2)  & 1.000 \\
O6    & 0&43468(6)  & 0&26748(2)  & 0&40966(3)  & 1.000 \\
O7    & 0&54949(5)  & $-$0&10620(1) & 0&38297(9)  & 1.000 \\
O8    & 0&91630(3)  & 0&06260(4)  & 0&08270(4)  & 1.000 \\
O9    & 0&32810(4)  & 0&16170(4)  & 0&01890(8)  & 1.000 \\
O10   & 0&14360(6)  & $-$0&28900(3) & 0&02900(2)  & 1.000 \\
D1    & 0&53894(2)  & 0&09933(6)  & 0&17231(7)  & 1.000 \\
D2    & 0&84020(3)  & 0&34348(1)  & 0&32517(8)  & 1.000 \\
D3    & 0&14204(6)  & 0&35221(3)  & 0&16796(8)  & 1.000 \\
D4    & 0&95022(6)  & 0&20807(1)  & 0&09791(9)  & 1.000 \\
D5    & 0&82915(2)  & 0&00218(3)  & 0&11198(4)  & 1.000 \\
D6    & 0&43125(3)  & 0&18719(7)  & $-$0&01040(2) & 1.000 \\
D7    & 0&35105(4)  & 0&25293(2)  & 0&05411(3)  & 1.000 \\
D8    & 0&24610(8)  & $-$0&27500(1) & 0&05808(3)  & 1.000 \\
D9    & 0&03547(2)  & $-$0&38630(6) & 0&04887(5)  & 1.000 \\ \hline\hline
\end{tabular}
\end{table}

\begin{table}[tb]
  \caption{\label{XRD_Stoe}Refined atomic positions in \ktH\ from powder x-ray diffraction using 1.540-\AA\ x-rays at 300\,K. The Zn1 site occupies the 2$a$ Wyckoff position, while all other sites occupy 4$e$.}
\begin{tabular}{lr@{.}lr@{.}lr@{.}lcc}\hline\hline
    Site & \multicolumn{2}{c}{$x$} & \multicolumn{2}{c}{$y$} & \multicolumn{2}{c}{$z$} & $U_\text{iso}$ & Occ. \\ \hline
Zn1  & 0&00000  & 0&00000  & 0&00000  & 0.012 & 1.00 \\
Cu1  & $-$0&00650(4) & 0&09700(3)  & 0&24831(2)  & 0.012 & 1.00 \\
Cu2  & 0&50220(6)  & $-$0&14470(1) & 0&25054(7)  & 0.012 & 0.363 \\
Zn2  & 0&50220(3)  & $-$0&14470(5) & 0&25054(8)  & 0.012 & 0.637 \\
S1   & 0&38475(2)  & 0&07586(4)  & 0&37486(8)  & 0.012 & 1.00 \\
O1   & 0&31435(7)  & 0&07693(3)  & 0&31408(11)  & 0.012 & 1.00 \\
O2   & 0&62845(5)  & 0&11817(2)  & 0&20860(2)  & 0.012 & 1.00 \\
O3   & 0&84728(5)  & 0&38203(3)  & 0&28635(3)  & 0.012 & 1.00 \\
O4   & 0&13652(4)  & 0&38021(4)  & 0&20858(6)  & 0.012 & 1.00 \\
O5   & 0&13911(3)  & 0&00931(8)  & 0&40270(4)  & 0.012 & 1.00 \\
O6   & 0&45780(4)  & 0&26517(5)  & 0&40062(3)  & 0.012 & 1.00 \\
O7   & 0&52687(7)  & $-$0&16180(2) & 0&39036(6)  & 0.012 & 1.00 \\
O8   & 0&94070(9)  & 0&05903(3)  & 0&08829(1)  & 0.012 & 1.00 \\
O9   & 0&32352(4)  & 0&18647(6)  & 0&01857(8)  & 0.012 & 1.00 \\
O10  & 0&14521(3)  & $-$0&32740(6) & 0&02581(5)  & 0.012 & 1.00 \\ 
\hline\hline
\end{tabular}
\end{table}

\sisetup{
  table-number-alignment = center,
  group-separator = {,},
  detect-weight = true,
  detect-family = true
}

\begin{table}[tb]
  \caption{\label{SCXRD_Ktenasite}Crystal data and structure refinement for ktenasite based on single-crystal x-ray diffraction.}
  \centering
  \setlength{\tabcolsep}{4pt} 
  \renewcommand{\arraystretch}{1.05}
  \begin{tabularx}{\columnwidth}{l >{\centering\arraybackslash}X}
    \hline\hline
    Empirical formula & \ktH \\
    Formula weight [g/mol] & \num{725.41} \\
    Temperature [K] & \num{180.0(2)} \\
    Crystal system & monoclinic \\
    Space group & $P2_1/c$ (no. 14) \\
    $a$ [\AA] & \num{5.5882(2)} \\
    $b$ [\AA] & \num{6.1730(2)} \\
    $c$ [\AA] & \num{23.6393(7)} \\
    $\beta$ [$^\circ$] & \num{95.570(3)} \\
    Volume [\AA$^3$] & \num{811.609(9)} \\
    $Z$ & \num{2.0} \\
    $\rho_{\mathrm{calc}}$ [g\,cm$^{-3}$] & \num{2.968(3)} \\
    $\mu$ [mm$^{-1}$] & \num{7.215(2)} \\
    $F(000)$ & \num{716.0} \\
    Crystal size [mm$^3$] & 0.008 $\times$ 0.050 $\times$ 0.114 \\
    Crystal colour & green \\
    Crystal shape & platelet \\
    Radiation & MoK$_{\alpha}$ ($\lambda$ = 0.71073 \AA) \\
    $2\theta$ range [$^\circ$] & 6.82--66.63 \\
    Index ranges & $-8 \le h \le 7$ \\
                 & $-8 \le k \le 9$ \\
                 & $-36 \le l \le 35$ \\
    Reflections collected & \num{7732.0} \\
    Independent reflections & \num{2704.0} \\
    $R_{\mathrm{int}}$ & \num{0.0207} \\
    $R_{\sigma}$ & \num{0.0256} \\
    Completeness to $\theta = 25.242^\circ$ & 98.4\% \\
    Data / Restraints / Parameters & 2704 / 0 / 139 \\
    Absorption correction $T_{\mathrm{min}}/T_{\mathrm{max}}$ & 0.7730 / 1.0000 \\
    Goodness-of-fit on $F^2$ & \num{1.064} \\
    \makecell[l]{Final $R$ indexes \\ $R_1/wR_1$ [I $\ge$ 2$\sigma$(I)]} & 0.0289 / 0.0714 \\
    \makecell[l]{Final $R$ indexes \\ $R_1/wR_1$ [all data]} & 0.0359 / 0.0737 \\
    Largest peak/hole [e\,\AA$^{-3}$] & 1.46 / $-$0.71 \\
    \hline\hline
  \end{tabularx}
\end{table}

\begin{table}[tb]
  \caption{\label{SCXRD_Ktenasite2}Refined atomic positions in \ktH\ at 180\,K from single-crystal x-ray diffraction using 0.7107-\AA\ x-rays. The Zn1 site occupies the 2$a$ Wyckoff position, while all other sites occupy 4$e$.}
\begin{tabular}{lr@{.}lr@{.}lr@{.}lcc}\hline\hline
    Site & \multicolumn{2}{c}{$x$} & \multicolumn{2}{c}{$y$} & \multicolumn{2}{c}{$z$} & $U_\text{iso}$ & Occ. \\ \hline
Zn1  & 0&00000  & 0&00000  & 0&00000  & 0.01087(10) & 1.00 \\
Cu1  & $-$0&01136(5) & 0&10218(5)  & 0&24884(2)  & 0.00787(8) & 1.00 \\
Cu2  & 0&49879(5)  & $-$0&14853(5) & 0&24663(2)  & 0.01143(8) & 0.37 \\
Zn2  & 0&49879(5)  & $-$0&14853(5) & 0&24663(2)  & 0.01143(8) & 0.63 \\
S1   & 0&36063(12)  & 0&05812(10)  & 0&37404(3)  & 0.01107(12) & 1.00 \\
O1   & 0&3307(4)  & 0&0968(3)  & 0&31167(8)  & 0.0133(4) & 1.00 \\
O2   & 0&6113(4)  & 0&1073(3)  & 0&20919(9)  & 0.0126(4) & 1.00 \\
O3   & 0&8405(3)  & 0&3581(3)  & 0&28993(9)  & 0.0120(3) & 1.00 \\
O4   & 0&1646(3)  & 0&3470(3)  & 0&214668(8)  & 0.0102(3) & 1.00 \\
O5   & 0&1273(4)  & $-$0&0076(3)  & 0&39279(9)  & 0.0183(4) & 1.00 \\
O6   & 0&4433(4)  & 0&2613(4)  & 0&40283(10)  & 0.0225(5) & 1.00 \\
O7   & 0&5413(4)  & $-$0&1134(3) & 0&38640(9)  & 0.0184(4) & 1.00 \\
O8   & 0&9182(4)  & 0&0604(3)  & 0&08404(8)  & 0.0161(4) & 1.00 \\
O9   & 0&3290(4)  & 0&1546(3)  & 0&01867(9)  & 0.0163(4) & 1.00 \\
O10  & 0&1421(4)  & $-$0&2968(3) & 0&02851(8)  & 0.0159(4) & 1.00 \\ 
H1   & 0&554(7)  & 0&124(6)  & 0&1767(18)  & 0.0220(10) & 1.00 \\
H2   & 0&857(8)  & 0&361(7)  & 0&322(2)  & 0.0320(12) & 1.00 \\
H3   & 0&161(8)  & 0&352(7)  & 0&1855(18)  & 0.0230(11) & 1.00 \\
H4   & 0&8794(1)  & 0&1964(2)  & 0&0874(2)  & 0.0240(0) & 1.00 \\
H5   & 0&7885(8)  & $-$0&0105(1)  & 0&0905(5)  & 0.0240(0) & 1.00 \\
H6   & 0&3649(4)  & 0&2273(4)  & $-$0&0110(4)  & 0.0240(0) & 1.00 \\
H7   & 0&3157(3)  & 0&2532(2) & 0&0446(4)  & 0.0240(0) & 1.00 \\
H8   & 0&2625(6)  & $-$0&2749(3)  & 0&0541(3)  & 0.0240(0) & 1.00 \\
H9   & 0&0367(4)  & $-$0&3656(2)  & 0&0465(1)  & 0.0240(0) & 1.00 \\
\hline\hline
\end{tabular}
\end{table}

\cleardoublepage
\bibliography{kteansite}
\clearpage
\section{Supplemental Material\label{supp}}
\setcounter{figure}{0}
\setcounter{table}{0}
\renewcommand{\thefigure}{S\arabic{figure}}
\renewcommand{\thetable}{S\arabic{table}}
\subsection{Scanning Electron Microscopy}
\begin{figure}[!htbp]
    \includegraphics[width=\columnwidth]{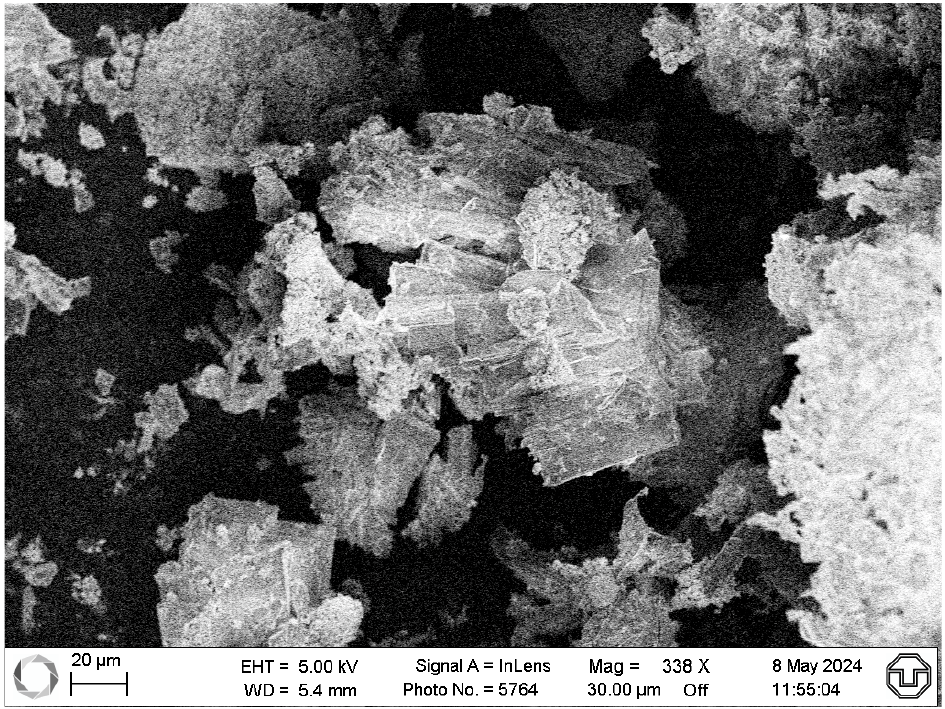}%\vspace{-12pt}
    \caption{High-resolution SEM micrograph of \ktH.}
    \label{FESEM}
\end{figure}

Surface morphology and elemental compositions were examined using SEM and EDS. The SEM image in Fig.~\ref{FESEM} reveals well-formed rectangular plate crystals ranging from 10--60\,$\mu$m, characteristic of monoclinic systems due to unequal lattice parameters.

\begin{figure}[!htbp]
    \includegraphics[width=\columnwidth]{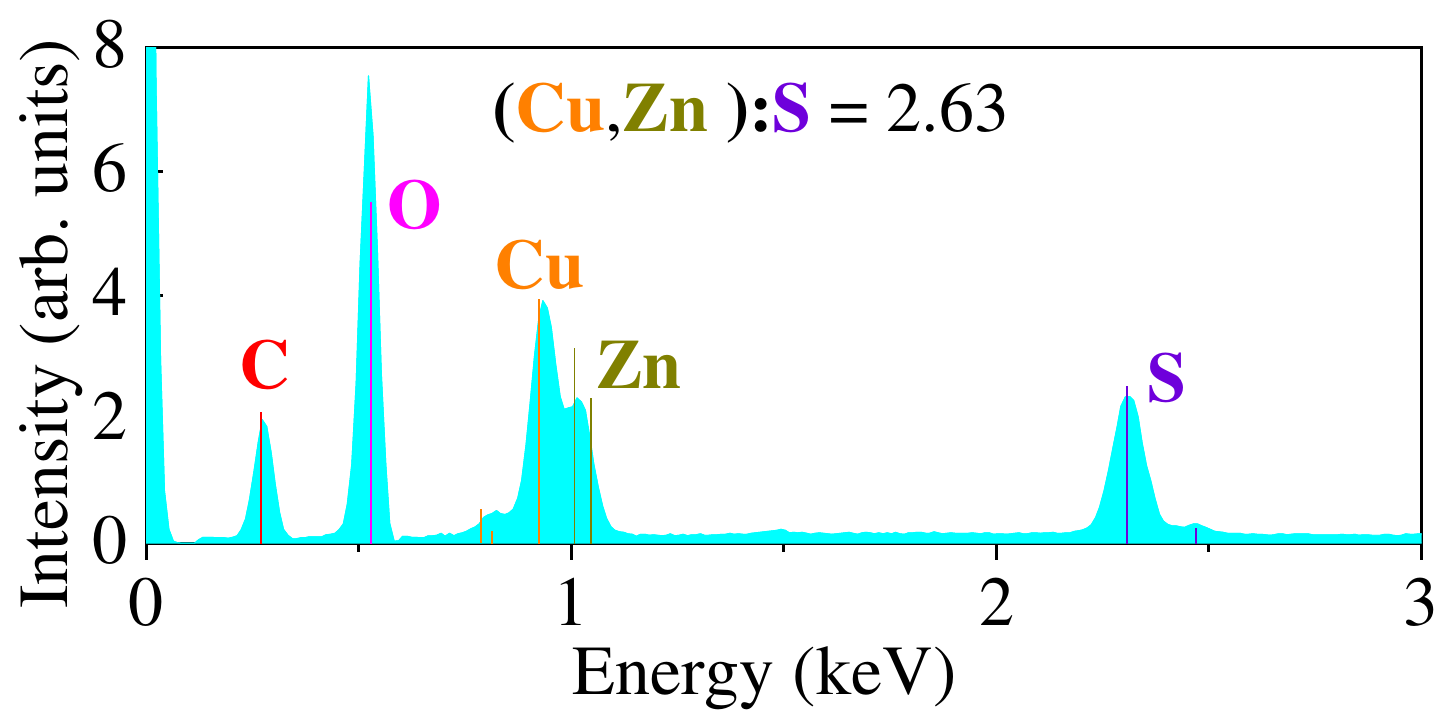}%\vspace{-10pt}
    \caption{EDS spectrum, illustrating the elemental composition of \ktH.}
    \label{EDX}
\end{figure}

The EDS spectrum in Fig.~\ref{EDX} confirms the presence of Cu, Zn, S, and O without detectable impurities. The detected carbon signal originates from the carbon tape used for mounting. The measured (Cu,Zn):S ratio of 2.63, closely matches the expected value of 2.50. However, distinguishing Cu and Zn by weight percentage using this technique is unreliable due to overlapping peaks, as evident in Fig.~\ref{EDX}.

\subsection{Thermal Analysis}
\begin{figure}[!htbp]
    \includegraphics[width=\columnwidth]{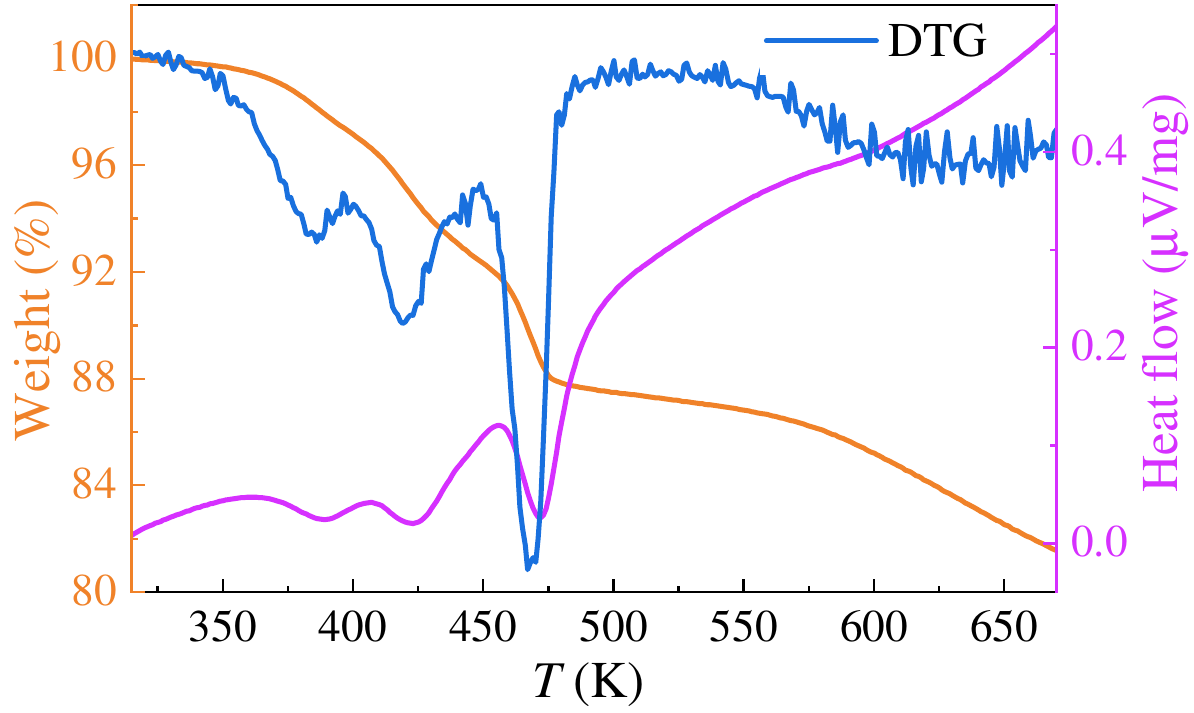}%\vspace{-1em}
    \caption{The TG, DTA and DTG curves of \ktH.}
    \label{TGA_DTA}
\end{figure}

Thermal analysis was performed to investigate phase stability and dehydration-driven phase transitions in \ktH. Figure~\ref{TGA_DTA} shows the thermogravimetric (TG), derivative thermogravimetric (DTG), and differential thermal analysis (DTA) curves. Minor mass losses at approximately 383\,K and 420\,K correspond to partial dehydration steps (approximately one and two H$_2$O groups, respectively). A sharper mass-loss event occurs near 470\,K, accompanied by a pronounced endothermic peak, by which point at least five of the six crystal waters have been released. The theoretical mass loss of 15\%\ when all six hydrate groups have been lost is not reached until significantly later, with no clear plateau, suggesting that there may already have been partial dehydration during storage at ambient conditions. Further decomposition is observed above 500\,K, continuing to temperatures beyond the measured range.

\subsection{Fourier-transform infrared spectroscopy}

\begin{figure}[!htbp]
    \includegraphics[width=\columnwidth]{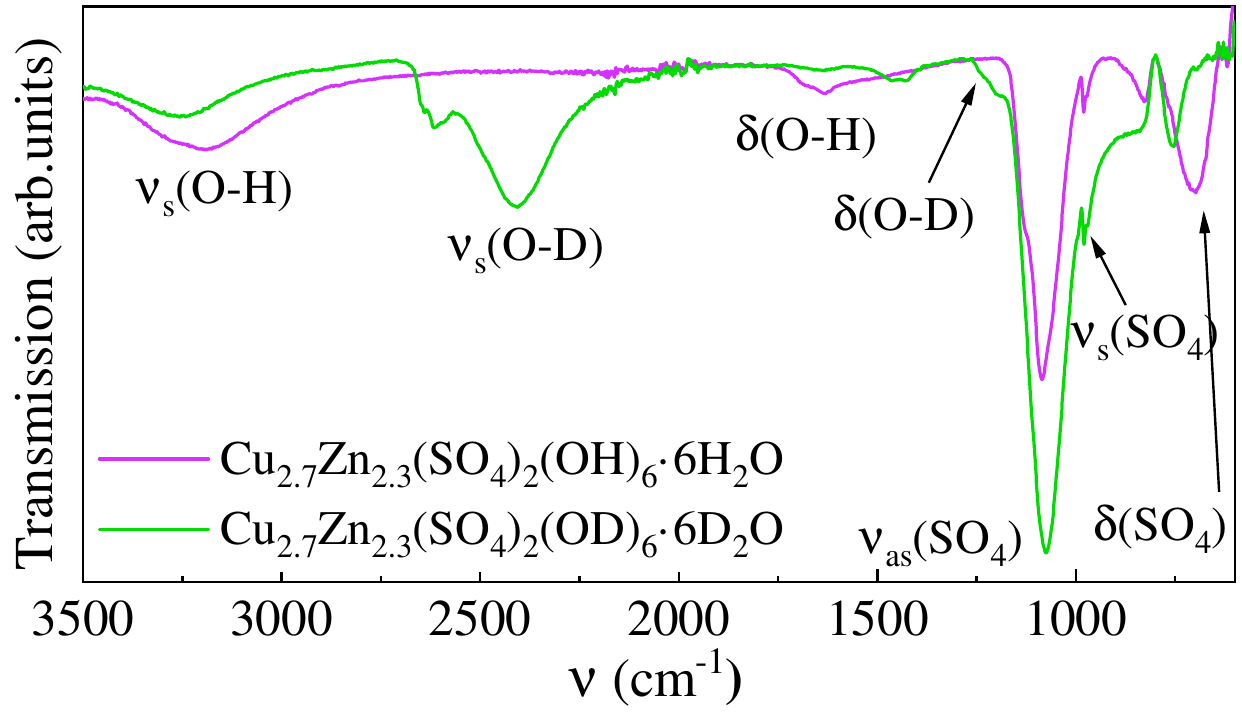}%\vspace{-1em}
    \caption{FTIR spectra of \ktH\ and \ktD\ at room temperature.}
    \label{FTIR}
\end{figure}

Fourier-transform infrared spectroscopy (FTIR) was employed to investigate bond characteristics and verify successful deuteration prior to performing neutron diffraction. The spectra of \ktH\ and \ktD\ in Fig.~\ref{FTIR} show a broad O--D stretching band near 2400\,cm$^{-1}$ in the deuterated compound which is absent in the protonated version, confirming significant H-D substitution. In \ktH, the O--H stretching and bending modes appear at 3200 and 1632\,cm$^{-1}$, respectively. The SO$_4^{2-}$ asymmetric and symmetric stretching modes are observed at 1075 and 975\,cm$^{-1}$, with the bending mode at 694\,cm$^{-1}$. A band at 830\,cm$^{-1}$ is attributed to metal-hydroxide bending vibrations.

\subsection{Neutron powder diffraction}
\begin{figure}[!htbp]
    \includegraphics[width=\columnwidth]{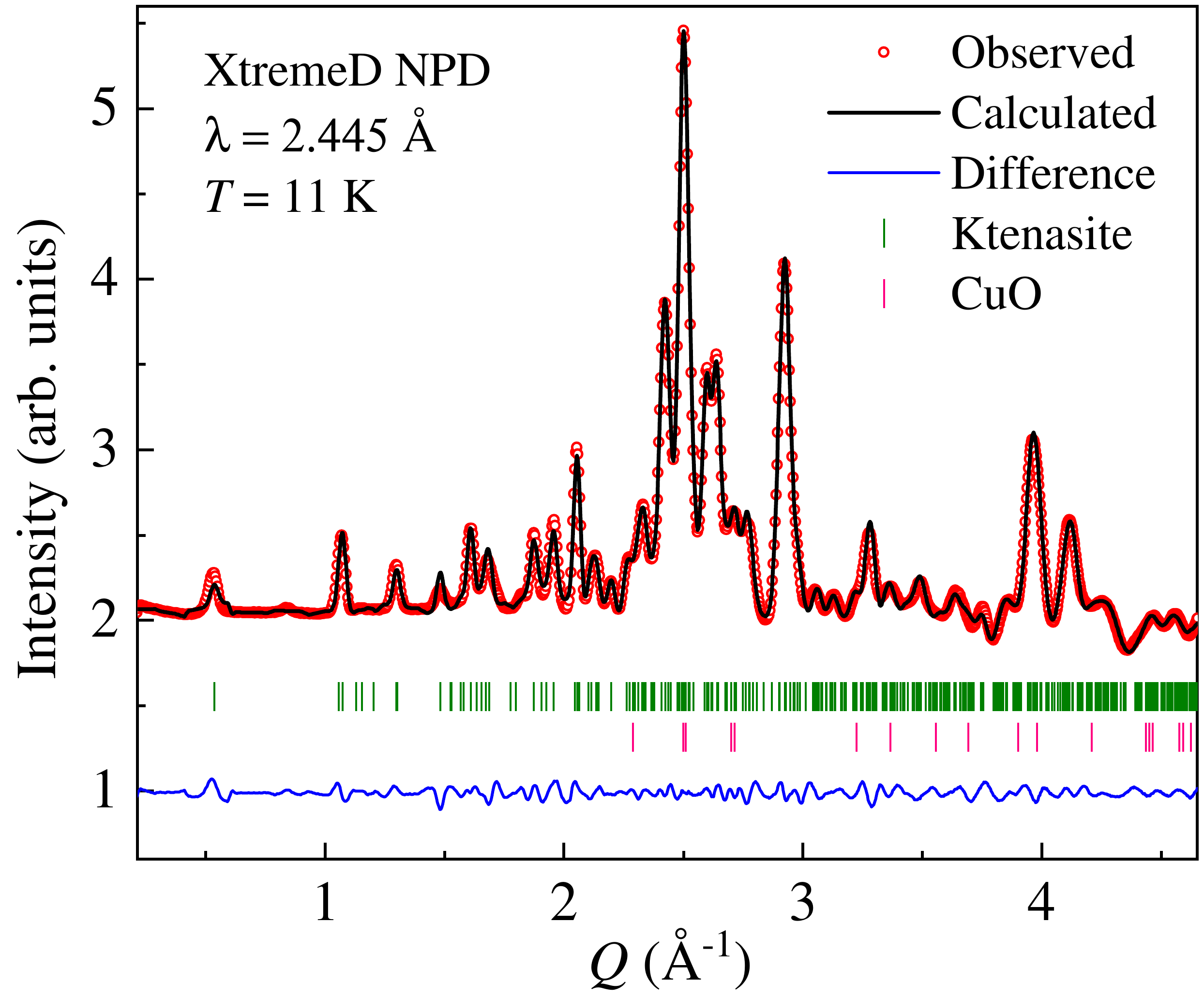}%\vspace{-1em}
    \caption{Rietveld-refined neutron powder diffraction pattern of \ktD\ at 11\,K.}
    \label{XtremeD}
\end{figure}

The neutron powder diffraction pattern collected at XtremeD using 2.445-\AA\ neutrons at 11\,K is shown in Fig.~\ref{XtremeD}, with the refined structural parameters and goodness-of-fit indicators summarized in Table~\ref{Summary_XtremeD} and the refined atomic positions in Table~\ref{NPD_XtremeD}.

\begin{table}[bt]
  \caption{\label{Summary_XtremeD}Summary of crystal structure refinement of \mbox{ktenasite} using NPD on XtremeD using 2.445-\AA\ neutrons.}
  \begin{ruledtabular}
    \begin{tabularx}{\columnwidth}{l@{\extracolsep{\fill}}c}
      Parameter & NPD \\ \hline
      Space group & $P2_1/c$ (no.\,14) \\
      $T$ (K) & 11 \\
      $a$ (\AA) & 5.5827(3) \\
      $b$ (\AA) & 6.1441(3) \\
      $c$ (\AA) & 23.5489(19) \\
      $\beta$ ($^\circ$) & 95.606(3) \\
      $V$ (\AA$^3$) & 803.88(9) \\
      $Z$ & 2 \\
      Density (g\,cm$^{-3}$) & 3.050(3) \\
      $Q$ range (\AA$^{-1}$) & 0.34--4.68 \\
      $R$ (\%) & 2.16 \\
      $wR$ (\%) & 1.51 \\
    \end{tabularx}
  \end{ruledtabular}
\end{table}

\begin{table}[!htbp]
  \caption{\label{NPD_XtremeD} Refined atomic positions in \ktD\ at 11\,K from neutron powder diffraction on XtremeD using 2.445-\AA\ neutrons. The Zn1 site occupies the 2$a$ Wyckoff position, while all other sites occupy 4$e$. The deuteration level refined to 76.60(2)\%.}
\begin{tabular}{l r@{.}l r@{.}l r@{.}l c}\hline\hline
Site & \multicolumn{2}{c}{$x$} & \multicolumn{2}{c}{$y$} & \multicolumn{2}{c}{$z$} & Occ. \\ \hline
Zn1  & 0&00000  & 0&00000  & 0&00000  & 1.000 \\
Cu1  & $-$0&02400 & 0&11400  & 0&24430  & 1.000 \\
Cu2  & 0&48900  & $-$0&12370 & 0&25210  & 0.360 \\
Zn2  & 0&48900  & $-$0&12370 & 0&25210  & 0.640 \\
S1   & 0&37100  & 0&07100  & 0&37570  & 1.000 \\
O1   & 0&36600  & 0&12500  & 0&31050  & 1.000 \\
O2   & 0&61500  & 0&12200  & 0&21240  & 1.000 \\
O3   & 0&81900  & 0&37600  & 0&28510  & 1.000 \\
O4   & 0&14600  & 0&34400  & 0&20860  & 1.000 \\
O5   & 0&14800  & 0&02800  & 0&39960  & 1.000 \\
O6   & 0&43500  & 0&26700  & 0&40970  & 1.000 \\
O7   & 0&54900  & $-$0&10600 & 0&38300  & 1.000 \\
O8   & 0&91630  & 0&06260  & 0&08270  & 1.000 \\
O9   & 0&32810  & 0&16170  & 0&01890  & 1.000 \\
O10  & 0&14360  & $-$0&28900 & 0&02900  & 1.000 \\
D1   & 0&57900  & 0&09700  & 0&17470  & 1.000 \\
D2   & 0&83500  & 0&34300  & 0&32280  & 1.000 \\
D3   & 0&08900  & 0&38300  & 0&17360  & 1.000 \\
D4   & 0&88600  & 0&17800  & 0&09760  & 1.000 \\
D5   & 0&83500  & $-$0&04300 & 0&10430  & 1.000 \\
D6   & 0&43100  & 0&20100  & $-$0&01390 & 1.000 \\
D7   & 0&38100  & 0&24800  & 0&05690  & 1.000 \\
D8   & 0&25900  & $-$0&28300 & 0&06000  & 1.000 \\
D9   & 0&04800  & $-$0&40800 & 0&05210  & 1.000 \\ \hline\hline
\end{tabular}
\end{table}

\subsection{Magnetic diffraction}
A preliminary analysis of the magnetic peaks confirms the incommensurate nature of the 
magnetic structure and the corresponding $k$-vector. Symmetry-allowed magnetic structures were explored via representation analysis with \textsc{SARA}\textit{h}, yielding the irreducible representation $\Gamma_1$. Our preliminary refinement, to be published separately once complete, yields a helical magnetic structure at 1.5\,K with an ordered moment of $\sim$\,$0.71\,\mu_{\text{B}}$.

\subsection{Crystallographic Information Files (CIFs)}

As ancillary files to this arXiv submission, we provide the following crystallographic information files (CIFs) describing our crystal structure refinements:

\begin{center}
\noindent\begin{tabular}{l}
\verb+D2B_10K_1p59A_Ktenasite.cif+\\
\verb+XtremeD_11K_2p45A_Ktenasite.cif+\\
\verb+Rigaku_180K_0p71A_Ktenasite.cif+\\
\verb+Stoe_300K_1p54A_Ktenasite.cif+\\
\end{tabular}
\end{center}

\end{document}